\documentclass[preprint,10pt]{elsarticle}

\usepackage{amssymb}

\usepackage{mathrsfs}
\usepackage[utf8]{inputenc}
\usepackage{amsmath}
\usepackage{amsfonts}
\usepackage{enumitem}  
\usepackage[ruled,vlined,linesnumbered,resetcount]{algorithm2e}
\usepackage{tikz}
\usetikzlibrary{positioning}
\usepackage{caption}
\usepackage{subcaption}

\usepackage{xcolor}
\usepackage{graphicx}
\usepackage[normalem]{ulem}
\usepackage{comment}

\usepackage{url}
\usepackage{hyperref}
\hypersetup{colorlinks=true, urlcolor=blue}

\usepackage{pgfplots}

\usepackage{bm}
\usepackage{nomencl}
\makenomenclature
\RequirePackage{doi}
\usetikzlibrary{arrows.meta}
\usetikzlibrary{angles, quotes}
\usetikzlibrary{calc}

\usepackage{minibox}
\usepackage[normalem]{ulem}
\usepackage[noend]{algpseudocode}

\usepackage{accents}

\usepackage{mathtools}

\journal{Computer Methods in Applied Mechanics and Engineering}

\newcommand{\txt}[1]{\textnormal{#1}}

\newcommand{\GlobalBasis}{\Psi}
\newcommand{\XBasis}[1]{\Phi_{#1}}
\newcommand{\TBasis}[1]{\mathcal{V}_{#1}}

\newcommand{\basis}[3]{\varphi_{#1_{#2#3}}}

\newcommand{\matL}[3]{X_{1_{#1#2}}^{#3}}

\newcommand{\matQconcat}{X_2}
\newcommand{\matLconcat}{X_1}

\newcommand{\param}[1]{\gamma_{#1}}

\newcommand{\lenspace}{\mathcal{N}}
\newcommand{\lensnap}{\overline{n}}
\newcommand{\lenparam}{\overline{m}}
\newcommand{\dimbasis}{\overline{q}}
\newcommand{\lenbasis}{(\overline{q}\cdot\overline{m})}

\newcommand{\SnapMat}[1]{\mathcal{U}_{#1}}

\newcommand{\tempvar}{v}

\newcommand{\usol}[2]{u_{#1}^{#2}}

\newcommand{\tempcoef}[3]{\tempvar_{#1_{#2}}^{#3}}
\newcommand{\tempcoefGILD}[3]{\tempvar_{*#1_{#2}}^{#3}}

\usepackage[a4paper, total={5.5in, 9.in}]{geometry}

\usepackage{floatrow}
\newfloatcommand{capbtabbox}{table}[][\FBwidth]

\SetKwInput{KwOffline}{Offline}
\SetKwInput{KwOnline}{Online}
\usepackage[export]{adjustbox}
\usetikzlibrary{positioning, fit, arrows.meta, shapes}
\begin{document}

\begin{frontmatter}



\author{R. Ayoub\fnref{label1}}
\fntext[label1]{rama.ayoub@kaust.edu.sa}
\author{M. Oulghelou\fnref{label2}}
\fntext[label2]{mourad.oulghelou@sorbonne-universite.fr}
\author{P. J. Schmid\fnref{label3}}
\fntext[label3]{peter.schmid@kaust.edu.sa}

\address{
$^{\txt{1,3}}$ Mechanical Engineering, King Abdullah University of Science and Technology, Thuwal, Saudi Arabia \\
$^{\txt{2}}$ Institute of Computing and Data Sciences, Sorbonne University, Paris, France\\
$^{\txt{2}}$ Jean Le Rond D'Alembert Institute, Sorbonne University, Paris, France}



\title{Improved Greedy Identification of Latent Dynamics with Application to Fluid Flows}


\begin{abstract}
Model reduction is a key technology for large-scale physical systems in science and engineering, as it brings behavior expressed in many degrees of freedom to a more manageable size that subsequently allows control, optimization, and analysis with multi-query algorithms. We introduce an enhanced regression technique tailored to uncover quadratic parametric reduced-order dynamical systems from data. Our method, termed Improved Greedy Identification of Latent Dynamics (I-GILD), refines the learning phase of the original GILD approach proposed in \cite{GILD2024}. This refinement is achieved by reorganizing the quadratic model coefficients, allowing the minimum-residual problem to be reformulated using the Frobenius norm. Consequently, the optimality conditions lead to a generalized Sylvester equation, which is efficiently solved using the conjugate gradient method. Analysis of the convergence shows that I-GILD achieves superior convergence for quadratic model coefficients compared to GILD's steepest gradient descent, reducing both computational complexity and iteration count. 
Additionally, we derive an error bound for the model predictions, offering insights into error growth in time and ensuring controlled accuracy as long as the magnitudes of initial error is small and learning residuals are well minimized. 
The efficacy of I-GILD is demonstrated through its application to numerical and experimental tests, specifically the flow past Ahmed body with a variable rear slant angle, and the lid-driven cylindrical cavity problem with variable Reynolds numbers, utilizing particle-image velocimetry (PIV) data. These tests confirm I-GILD's ability to treat real-world dynamical system challenges and produce effective reduced-order models.
\end{abstract}


\end{frontmatter}
\newpage

\section{Introduction}
{
Model reduction techniques have become essential in fields where solving high-dimensional systems or handling large datasets poses significant computational challenges. These techniques reduce computational costs by preserving the core features and dynamics of complex systems in a lower-dimensional representation, enabling applications in flow control, design optimization, uncertainty quantification, and multi-query algorithms such as Monte Carlo simulations.
\\
A core aspect of model reduction involves data-compression methods, which produce low-dimensional representations of high-dimensional data. Proper Orthogonal Decomposition (POD)~\cite{Sirovich} is a widely used \emph{linear} technique for this purpose, identifying orthonormal modes that capture dominant structures and significant variations in the data.
For \emph{nonlinear} model reduction, auto-encoder neural networks~\cite{SCHMIDHUBER201585} and t-SNE (t-distributed Stochastic Neighbor Embedding)~\cite{maaten2008visualizing} can effectively map high-dimensional data into a low-dimensional latent space. In this paper, we rely on POD for its advantageous balance between computational simplicity and interpretability.
\\
Beyond data compression, predicting the latent dynamics of reduced variables is crucial for understanding the evolution of complex systems. Intrusive methods, such as Galerkin projection of the Navier--Stokes equations, can be accurate but require full knowledge of the governing equations, thus limiting their applicability in black-box or partially observed scenarios. Consequently, \emph{non-intrusive}, data-driven approaches that bypass Galerkin projection have gained popularity. For instance, Dynamic Mode Decomposition (DMD)~\cite{Schmid2008DynamicMD,DMD2022AnRFM,DUAN2024112621}---closely connected to the concept of Koopman operator---extracts the dominant dynamic modes of a system, while spectral-submanifold methods~\cite{Cenedese2022HallerSSM,haller2022SSM} enable predictive modeling in a reduced phase space. Recent advances also incorporate machine learning and stochastic ideas: Sparse Identification of Nonlinear Dynamics (SINDy)~\cite{brunton_kutz_2019,brunton2016discovering,champion2020unified} relies on sparse regression to discover governing equations from flexible functional libraries; generative approaches forecast the evolution of high-dimensional flows~\cite{Seb2024generative}; stochastic models address latent-space uncertainties~\cite{pan2023latent_uncertainty}. 
Meanwhile, deep-learning strategies capture nonlinear temporal dependencies in fluid systems~\cite{renganathan2020machine,maulik2020time}. These can be used in concert with self-attention and auto-encoder architectures~\cite{fu2021autoencoder}. 
Physics-informed approaches, including greedy latent space dynamics identification (gLaSDI) framework \cite{marion2024glasi}, and
Physics-Informed Neural Networks (PINNs)~\cite{raissi2019physics,mendez2023machine}, together with operator-based approaches~\cite{chen2018neural,li2020fourier,Linot2022}, incorporate PDE constraints or parametric mappings—though they may require extensive data and careful tuning. Compared to these broader strategies, polynomial-based models seem to be more suitable in the identification of POD latent dynamics, balancing computational efficiency with physical interpretability. In particular, quadratic models follow the truncated Navier--Stokes Galerkin ROM framework.
\\
In recent years, a growing body of work has emphasized the effectiveness of \emph{polynomial}---most notably, \emph{quadratic}---representations in data-driven reduced-order models (ROMs). By providing a flexible yet interpretable framework, polynomial expansions can accurately capture the nonlinear interactions often lost in purely linear approaches. For instance, advanced manifold-learning techniques leverage polynomial-form embeddings to learn physics-based reduced models~\cite{kim2024nonlinear_manifolds}, while non-intrusive, optimization-driven formulations jointly determine projection operators and polynomial reduced dynamics~\cite{Padovan2024}. In fluid-flow modeling, polynomial closures offer a data-driven mechanism to approximate unresolved mode interactions and energy transfers; examples include \emph{quadratic} ansatz methods that filter out high-frequency flow components \cite{Xie2018}. Operator-inference approaches similarly exploit polynomial terms to approximate unknown nonlinear components~\cite{PEHERSTORFER2016196}. Even DMD frameworks have been upgraded with \emph{quadratic manifolds} to minimize projection errors and capture rich flow dynamics~\cite{Kalur2023}. Altogether, these polynomial-based schemes strike a practical balance: they extend linear ROMs to handle key nonlinearities without resorting to large-scale neural networks or intrusive PDE closures, thus reinforcing both accuracy and interpretability. 
In this context, the \emph{Greedy Identification of Latent Dynamics} (GILD) approach proposed in~\cite{GILD2024} leverages a multivariate quadratic ansatz to model reduced-order dynamics. This formulation, inspired by the truncated Navier--Stokes Galerkin ROM framework, incorporates parameter variation through interpolation across reduced subspaces~\cite{Oulghelou_Quotient_manifold_interpolation}. These subspaces are defined within the quotient manifold of the set of maximal rank matrices by the orthogonal group, enabling efficient interpolation while preserving the mathematical form of the model. The straightforward structure of GILD eliminates the need for complex neural architectures or extensive hyperparameter tuning, while remaining sufficiently expressive to capture high-dimensional fluid dynamics.
\\
Despite its effectiveness, GILD’s reliance on a steepest-descent algorithm for residual minimization proves computationally expensive, particularly for large-scale or parametric systems. To address these challenges, we propose an enhanced framework called \emph{Improved Greedy Identification of Latent Dynamics} (I-GILD). The learning phase of I-GILD reorganizes the residuals into a matrix form, allowing the minimization problem to be reformulated under a Frobenius-norm objective. This reformulation naturally leads to a \emph{generalized Sylvester equation} of the form
$$\sum_m \mathbf{A}_m\,\mathbf{X}\,\mathbf{B}_m = \mathbf{F},$$
where $\mathbf{X}$ encodes the sought polynomial operator, and $\mathbf{A}_m,\mathbf{B}_m,\mathbf{F}$ contain the data-driven residual terms. The \emph{key improvement} over steepest descent is the use of low-rank approximate solutions to the Sylvester equations~\cite{JBILOU2006365}. In our case, we employ a \emph{conjugate gradient (CG)} solver~\cite{sym14091868,CHEN20229925}, which efficiently exploits the symmetric structure of the Sylvester matrix equation in I-GILD framework to converge faster. By updating $\mathbf{X}$ along \emph{conjugate} directions, I-GILD substantially reduces the iterations required to minimize the residual.
\\
To gain insight into how the model is expected to evolve over time a priori, we derive an \emph{error bound} for the training samples. This bound illustrates when short-term error growth remains controlled under bounded spectral norms of the identified operators, moderate initial errors and well-minimized residuals. Numerical experiments on the Ahmed body flow and a lid-driven cylindrical cavity demonstrate that I-GILD not only converges more rapidly but also scales effectively to large or parametric datasets. These results highlight I-GILD’s potential for a wide range of complex, real-world flows.
\\
The remainder of this paper is organized as follows: Section~2 reviews the original GILD method~\cite{GILD2024}. Section~3 presents the I-GILD framework and discusses its computational efficiency and a priori error analysis. Numerical experiments on Ahmed body flows with varying rear slant angle, as well as lid-driven cavity flows at different Reynolds numbers, are provided in Section~4. Finally, Section~5 summarizes the key contributions of this work and suggests avenues for future research.
}

\section{Review of GILD}

We start by presenting an overview of the Greedy Identification of Latent Dynamics (GILD) method \cite{GILD2024}, which is designed to extract latent dynamics from parametric flow data. This part will set the stage for developing an improved algorithm in the subsequent section. 

Consider a set of $\lenparam$ parameter values denoted by $\{\param{m}\}_{0,\dots,\lenparam}$, and a corresponding set of $\lensnap+1$ snapshots of the system's state at a given parameter value $\param{m}$, represented by $\{u(t_n,x,\param{m}) := \usol{m}{n}\}_{n=0,\dots,\lensnap}$. 
Here, $x \in \Omega \subset \mathbb{R}^{d}$ denotes the $d$ spatial coordinates, and $t_n = n \, \Delta t$ represents the $n$-th discrete time instants. The snapshots, uniformly sampled over a time window of length $T$, are taken at intervals of size $\Delta t$.

Each snapshot captures the system's state, such as a two- or three-dimensional fluid velocity field, at $\lenspace$ spatial locations in $\Omega$. These fields are reshaped into high-dimensional column vectors in $\mathbb{R}^{d\, \lenspace}$, and organized into the following snapshot matrix for each parameter:

\begin{equation*}
    \SnapMat{m} = 
    \begin{bmatrix}
    | & | & & | \\
    u(t_0,\cdot, \param{m}) & u(t_1,\cdot, \param{m}) & \cdots & u(t_{\lensnap},\cdot, \param{m}) \\
    | & | & & |
    \end{bmatrix}, \quad m=1,\dots,\lenparam
\end{equation*}
To reduce the dimensionality of this snapshot matrix, a low-rank basis $\XBasis{m}{}$ is sought. This basis can be obtained by solving a projection problem that minimizes the residual between the original data and its projection onto a lower-dimensional subspace. 

{
Specifically, for $m=1,\dots,\lenparam$, the projection problems can be formulated as
\begin{equation*}
    \underset{\bm{\alpha_m}, \XBasis{m}{}}{\min} \,\sum_{n=0}^{\lensnap}\, \left\| \usol{m}{n} - \XBasis{m}{} \alpha^n_m\right\|^2_{\mathscr{A}}, \qquad \text{subject to} \quad \XBasis{m}{}^T\mathscr{A}\XBasis{m}{} = I_{\dimbasis},
\end{equation*}
where $\alpha_m^n \in \mathbb{R}^{\dimbasis}$ represents the vector of temporal coefficients at step $n$, $\|\cdot\|_{\mathscr{A}}$ denotes the norm induced by the inner product associated with the symmetric positive-definite weighting matrix $\mathscr{A}$ (commonly the mass matrix), and $I_{\dimbasis}$ is the $\dimbasis \times \dimbasis$ identity matrix, with $\dimbasis$ corresponding to the reduced basis dimension, often determined by the truncation order in the POD procedure. The orthogonality constraint $\XBasis{m}{}^{T}\mathscr{A}\XBasis{m}{} = I_{\dimbasis}$ ensures that the basis functions form an $\mathscr{A}$-orthonormal set. This formulation determines the best linear approximation of the snapshot data in a reduced subspace of dimension $\dimbasis$, significantly smaller than the original high-dimensional space, thereby facilitating computational efficiency while retaining essential dynamical features.
}

Using the snapshot method \cite{Sirovich}, the reduced basis $\XBasis{m}{}$ can be computed as $\XBasis{m}{} = \mathcal{U}_m \, \TBasis{m} \, \Sigma_m^{-1}$, where $\TBasis{m}$ arises from the spectral decomposition $(\mathcal{U}_m^T \mathscr{A} \mathcal{U}_m) \, \TBasis{m} = \TBasis{m} \Sigma_m^2$. The diagonal matrix $\Sigma_m$ contains the square roots of the POD eigenvalues, which represent the energy distribution across the modes. Thus, using POD, the snapshots can be approximated in the reduced subspaces $\text{Span}\{\XBasis{m}\}$, with dimension $\dimbasis \ll \lensnap$, as

\begin{equation}\label{POD_niveau1}
    \SnapMat{m} \, \approx \, \XBasis{m} \, \Sigma_m \, \TBasis{m}^T, \quad m=1,\dots,\lenparam
\end{equation}
where $\TBasis{m}$ encodes the temporal dynamics, and its rows $\tempcoef{m}{}{n^T}$ capture the latent time evolution at discrete time instant $t_n$, such that:

\begin{equation}\label{expression_of_V_and_V_dot}
    \TBasis{m} =
    \begin{bmatrix}
    \tempcoef{m}{}{0^T} \\
    | \\
    \tempcoef{m}{}{\lensnap^T}
    \end{bmatrix} =
    \begin{bmatrix}
    \tempcoef{m}{,1}{0} & \tempcoef{m}{,2}{0} & \cdots & \tempcoef{m}{,\dimbasis}{0} \\
    | & | & & | \\
    \tempcoef{m}{,1}{\lensnap} & \tempcoef{m}{,2}{\lensnap} & \cdots & \tempcoef{m}{,\dimbasis}{\lensnap}
    \end{bmatrix}, \quad m=1,\dots,\lenparam
\end{equation}
Next, in order to delineate the parametric dependency from the spatial direction, a second POD is performed as follows:
\begin{equation*}
    \begin{bmatrix}
    | & | & & | \\
    \XBasis{1} \Sigma_1 & \XBasis{2} \Sigma_2 & \cdots & \XBasis{\lenparam} \Sigma_{\lenparam} \\
    | & | & & |
    \end{bmatrix}
    = \GlobalBasis \, \Theta \, \begin{bmatrix}     \varphi_1 & \varphi_2 & \cdots &    \varphi_{\lenparam} \end{bmatrix},
\end{equation*}
where $\GlobalBasis$ is the reduced basis spanning the entire parametric snapshot space, and $\varphi_m$ contains the parametric dependency. The reduced bases $\varphi_m$ verify $\varphi_m^T  \varphi_{m'} = \delta_{m m'} \, I_{\dimbasis}$ with $\delta_{m m'}$ the Kronecker index.
GILD suggests to express the data-driven parametric reduced-order model as a quadratic multivariate polynomial in terms of the latent variables $\tempcoef{m}{}{n}$ and the reduced projection basis $\basis{m}{}{}$. The reduced dynamical system coefficents are sought such that the magnitude of the residuals $\bm{r}_m^n \in \mathbb{R}^{\dimbasis}$, for parameter $\param{m}$ and time step $t_n$, are greedily minimized, where
\begin{equation*}
    \bm{r}_m^n = \basis{m}{}{}^T \basis{m}{}{} \, \tempcoef{m}{}{n} - \basis{m}{}{}^T \, \matLconcat \, \basis{m}{}{} \, \tempcoef{m}{}{n-1} - \,   \basis{m}{}{}^T \, \matQconcat \, B_m^{n-1} \, \tempcoef{m}{}{n-1},
\end{equation*}
simplifying to 
\begin{equation*}
    \bm{r}_m^n =  \tempcoef{m}{}{n} - \basis{m}{}{}^T \, \matLconcat \, \basis{m}{}{} \, \tempcoef{m}{}{n-1} - \,   \basis{m}{}{}^T \, \matQconcat \, B_m^{n-1} \, \tempcoef{m}{}{n-1},
\end{equation*}
where $B_m^{n-1} = (\basis{m}{}{}\tempcoef{m}{}{n-1}) \otimes \basis{m}{}{}$ is defined using the Kronecker product. Here, $\matL{}{}{} \in \mathbb{R}^{\lenbasis\times\lenbasis}$ and $\matQconcat \in \mathbb{R}^{\lenbasis\times\lenbasis^2}$ are, respectively, the linear and quadratic coefficients of the model.
Note that, unlike the model introduced in \cite{GILD2024} which includes a bias term, we consider here an unbiased model. The inclusion of this term in the quadratic model is deemed unnecessary due to the fact that $\tempcoef{m}{}{n}$ lie within the unit ball in $\mathbb{R}^{\dimbasis},$ meaning that $\|\tempcoef{m}{}{n}\| \leq 1$. Introducing a bias term could potentially shift the resulting vector $\tempcoef{m}{}{n}$ outside this unit ball, thereby violating the norm constraint that naturally arises from the structure of of the latent vectors. Consequently, the absence of a bias term ensures that the model remains translation-invariant, with the transition from $\tempcoef{m}{}{n-1}$ to $\tempcoef{m}{}{n}$ being governed purely by the quadratic mapping, without any external offset, in accordance with the geometric constraints imposed by the unit ball.

GILD operates by progressively minimizing the residuals across all snapshots, following a two-stage order-greedy strategy: starting from order one (linear) and completing by order two terms (quadratic). Specifically, the minimization problems $\mathcal{P}_k$ for $k=1,2$ are expressed as
\begin{align}\label{GILD_min_pb_2norm}
    \mathcal{P}_{_k}: \quad & \underset{X}{\min} \,\, \mathcal{J}_k(X), \qquad \text{with} \qquad \mathcal{J}_k(X) = \frac{1}{2} \sum_{m=1}^{\lenparam} \sum_{n=1}^{\lensnap} \, \left\| r_{m}^{k,n}(X) \right\|^2 + \frac{\omega}{2} \, \left\| X \right\|^2.
\end{align}
The constant $\omega>0$ is the Tikhonov regularization parameter, introduced to prevent over-fitting. It controls the trade-off between fitting the data closely and penalizing large model weights, thus enhancing the model's stability and ability to generalize to unseen data. The terms $r_{m}^{k,n}$ are the residuals given, for $n = 1, \dots, \lensnap$, by:
\begin{align*}
    r_{m}^{1,n}&= \tempcoef{m}{}{n}  - \basis{m}{}{}^T X \basis{m}{}{} \tempcoef{m}{}{n-1}, \quad X  \in \mathbb{R}^{\lenbasis\times\lenbasis}, \\
    r_{m}^{2,n} &=  r_{m}^{1,n} - \basis{m}{}{}^T X B_{m}^{n-1} \tempcoef{m}{}{n-1}, \quad X  \in \mathbb{R}^{\lenbasis\times\lenbasis^2}.
\end{align*}
The necessary condition for optimality in the minimization problems $\mathcal{P}_{_k}$ is that the gradient of the cost functional $\mathcal{J}_k(X)$ with respect to $X$ vanishes at the minimum, i.e., $\nabla_{_X} \mathcal{J}_k = 0$. Solving this problem leads to a system of linear equations for each $\mathcal{P}_k$. As noted in \cite{GILD2024}, solving $\mathcal{P}_2$ has been identified as the most computationally intensive problem, due to the dense nature of the Gram matrix and its large size of $\lenbasis^3 \times \lenbasis^3$, which renders storage and computations impractical. Therefore, this challenge was addressed in GILD by employing the iterative steepest-descent algorithm.

In terms of a generalization to unseen parameters, GILD's model construction allow it to adapt with respect to parameter change by using an interpolation strategy of the subspaces $\txt{span}(\basis{m}{}{})$. This interpolation is performed in the quotient manifold $\mathbb{R}^{\lenbasis \times \dimbasis}_{*}/\mathcal{O}_{\dimbasis}$, as detailed in \cite{Oulghelou_Quotient_manifold_interpolation,GILD2024}. For a new parameter $\param{*}$, it consists in iterating the  following sequence
\begin{equation}\label{interp_phi_*}
    \varphi_{*}^{(h+1)} = \sum_{m=1}^{\lenparam} \omega_m(\param{*}) \, \varphi_m Q_{* m}^{(h)}, \quad h\geq 0
\end{equation}
where $\omega_m$ are the interpolation weights and $Q_{* m}^{(h)} = V_{* m}^{(h)} U_{* m}^{(h)^T}$ are orthogonal matrices formed from the left and right singular matrices of $\varphi_*^{(h)^T} \Theta^2 \varphi_m$. The converged solution $\varphi_{*}$ provides the barycenter for the new parameter $\param{*}$, and the approximated velocity field for $\param{*}$ at step $n$ can then be reconstructed in the high-dimensional space as
\begin{equation}\label{approx u* udot*}
    u_*^{n} = \GlobalBasis \, \Theta \, \basis{*}{}{} \, \tempcoef{*}{}{n}.
\end{equation}
where, by setting $B_*^{n-1} = (\basis{*}{}{}\tempcoef{*}{}{n-1}) \otimes \basis{*}{}{}$, the latent vector $\tempcoef{*}{}{n}$ was predicted by solving the adapted reduced-order model
\begin{equation}\label{Eq_Gild_nu*}
    \basis{*}{}{}^T \basis{*}{}{} \tempcoef{*}{}{n} = \basis{*}{}{}^T \, \matLconcat \, \basis{*}{}{} \, \tempcoef{*}{}{n-1} + \basis{*}{}{}^T \, \matQconcat \, B_*^{n-1} \, \tempcoef{*}{}{n-1}. 
\end{equation}

In summary, the GILD method provides a comprehensive framework for deriving reduced-order models by progressively minimizing residuals across time and parameter sampling points. However, it should be noted that the computational intensity of the method, particularly in solving the quadratic model coefficients, presents a significant limitation in terms of convergence speed and scalability.
Therefore, in the next section, we address these challenges by proposing an improvement to the GILD learning phase. Specifically, we reformulate the minimization problems \eqref{GILD_min_pb_2norm} using the Frobenius norm, in place of the $2$-norm.  This allows an intuitive use of the conjugate-gradient method, instead of the steepest-descent algorithm, accelerates convergence, and improves the overall computational efficiency of the original method.

\section{Improved GILD}

In this section, we present an enhanced formulation of the GILD model learning, addressing the computational challenges encountered in the original approach. To this end, we begin by introducing the matrix concatenating the residuals 

\begin{equation*}
    \mathcal{R}_{m,k}
    = 
    \begin{bmatrix}
    r_m^{1,k^T}  \\
    r_m^{2,k^T}  \\
    | \\
    r_m^{\lensnap,k^T} 
    \end{bmatrix}
    =
    \begin{bmatrix}
    r_{m_{,1}}^{1,k} & r_{m_{,2}}^{1,k} & \cdots & r_{m_{,\dimbasis}}^{1,k} \\
    r_{m_{,1}}^{2,k} & r_{m_{,2}}^{2,k} & \cdots & r_{m_{,\dimbasis}}^{2,k} \\
    | & | & & | \\
    r_{m_{,1}}^{\lensnap,k} & r_{m_{,2}}^{\lensnap,k} & \cdots & r_{m_{,\dimbasis}}^{\lensnap,k}
    \end{bmatrix} \in \mathbb{R}^{\lensnap \times \dimbasis}, 
\qquad
k= 1, 2.
\end{equation*}
The primary objective is to express $\mathcal{R}_{m,k}$ in terms of the multivariate polynomial  model coefficients $X_1$ and $X_2$. To do so, we introduce the temporal matrices $V_m$ and their forward-shifted counterparts $W_m$, which are structured as follows

\begin{equation}\label{expression_of_V_and_V_dot}
    V_m = 
    \begin{bmatrix}
    \tempcoef{m}{,1}{0} & \tempcoef{m}{,2}{0} & \cdots & \tempcoef{m}{,\dimbasis}{0} \\
    \tempcoef{m}{,1}{1} & \tempcoef{m}{,2}{1} & \cdots & \tempcoef{m}{,\dimbasis}{1} \\
    | & | & & | \\
    \tempcoef{m}{,1}{\lensnap-1} & \tempcoef{m}{,2}{\lensnap-1} & \cdots & \tempcoef{m}{,\dimbasis}{\lensnap-1}
    \end{bmatrix},  
    \qquad
    W_m = 
    \begin{bmatrix}
    \tempcoef{m}{,1}{1} & \tempcoef{m}{,2}{1} & \cdots & \tempcoef{m}{,\dimbasis}{1} \\
    \tempcoef{m}{,1}{2} & \tempcoef{m}{,2}{2} & \cdots & \tempcoef{m}{,\dimbasis}{2} \\
    | & | & & | \\
    \tempcoef{m}{,1}{\lensnap} & \tempcoef{m}{,2}{\lensnap} & \cdots & \tempcoef{m}{,\dimbasis}{\lensnap}
    \end{bmatrix}.
\end{equation}
Next, we define the matrices $\Psi_{1,m}\in\mathbb{R}^{\lenbasis\times \lenbasis}$ and $\Psi_{2,m}\in\mathbb{R}^{\lenbasis\times \lenbasis^2}$, which play a key role in achieving the sought-after expressions for $X_1$ and $X_2.$ Specifically, these matrices are given by
\begin{equation}\label{Psi_k equations}
    \Psi_{1,m} = \basis{m}{}{} V_m^T,
    \qquad
    \textrm{and}
    \qquad
    \Psi_{2,m} = \left( \basis{m}{}{} V_m^T \otimes \basis{m}{}{} \right) \left( \sum_{n=0}^{\lensnap-1} \delta_n \otimes \tempcoef{m}{}{n} \right)
\end{equation}
where $\delta_n$ is the $\lensnap$-by-$\lensnap$ Kronecker matrix, taking on the value $1$ for the $n$-th diagonal element and zero otherwise. Using the expressions for $\Psi_{1,m}$ and $\Psi_{2,m}$, we can now express the matrix of residuals $ \mathcal{R}_{m_{,k}}$ as follows

\begin{align*}
    \mathcal{R}_{m_{,1}} &= W_m - \Psi_{1,m}^T X_1^T \basis{m}{}{}, \quad X_1 \in \mathbb{R}^{\lenbasis\times\lenbasis} \\
    \mathcal{R}_{m_{,2}} &= \mathcal{R}_{m_{,1}} - \Psi_{2,m}^T X_2^T \basis{m}{}{}, \quad X_2 \in \mathbb{R}^{\lenbasis\times\lenbasis^2}.
\end{align*}
This matrix form of the residuals allows us to formulate the problem of identification of model's coefficients by minimizing the residuals $\mathcal{R}_{m_{,k}}$ using the Frobenius norm. Thus, in the I-GILD framework, the problems to solve for $k=1,2$ are expressed as

\begin{align*}
    \widetilde{\mathcal{P}}_{_k} \quad &: \quad \underset{X}{\min} \,\, \mathcal{J}_k(X), \quad \text{with} \quad \mathcal{J}_k(X) = \frac{1}{2} \sum_{m=1}^{\lenparam} \, \left\| \mathcal{R}_{m_{,k}} \right\|_F^2 + \frac{\omega}{2} \, \left\| X \right\|_F^2.
\end{align*}
To find the optimal coefficients $X$, we derive the necessary conditions by setting $\nabla_{_X}\mathcal{J}_k$ to zero. The optimality conditions for $\widetilde{\mathcal{P}}_{_k}$ lead to the following system of equations
\begin{equation}\label{Sylv_prob}
    \sum_{m=1}^{\lenparam} A_m X_k D_{k,m} - G_k + \omega X_k = 0, \quad\quad k=1,2
\end{equation}
where
\begin{equation*}
    G_1 = \sum_{m=1}^{\lenparam} \basis{m}{}{} W_{m}^T \Psi_{1,m}^T, \quad G_2 = \sum_{m=1}^{\lenparam} \basis{m}{}{} \mathcal{R}_{m_{,1}}^T \Psi_{2,m}^T, \quad A_m = \basis{m}{}{} \basis{m}{}{}^T, \quad D_{k,m} = \Psi_{k,m} \Psi_{k,m}^T,
\end{equation*}
By setting $A_0 = \sqrt{\omega} \, I_{\lenbasis}$ and $D_{k,0} = \sqrt{\omega} \, I_{\lenbasis}$, the system \eqref{Sylv_prob} reduces to the following generalized Sylvester problem

\begin{equation}\label{MLROM Sylvester_problem}
    \sum_{m=0}^{\lenparam} A_m X_k D_{k,m} = G_k, \qquad k=1,2.
\end{equation}
A straightforward approach to solving the generalized Sylvester equations above is to  vectorize the matrix equation using the Kronecker product. To this end, we introduce 
\begin{equation}\label{Mk_kron}
    \mathcal{M}_{k} = \sum_{m=0}^{\lenparam} D_{k,m}^T \otimes A_m, \qquad k=1,2
\end{equation}
and convert the problem into a large system of linear equations
\begin{equation}\label{linear system Sylvester Pb}
    \mathcal{M}_{k} \text{vec}(X_k) = \text{vec}(G_k).
\end{equation}
Given that $\mathcal{M}_{k}$ is SPD (symmetric positive definite), and thus non-singular, a solution $X_k$ to problem \eqref{linear system Sylvester Pb} exists and is unique.
To solve this system, one can apply the conjugate-gradient (CG) method, a powerful iterative algorithm designed for SPD systems. The CG method constructs a sequence of increasingly accurate solutions in the Krylov subspaces $\mathcal{K}_{k,J}$ of dimension $J$. Starting from a zero initialization of the solution $X_k^{(0)}=0$, the Krylov subspace is given by  
\begin{equation*}
    \mathcal{K}_{k,J}(\mathcal{M}_{k}, \text{vec}(G_k) ) = \text{span} \{\text{vec}(G_k), \mathcal{M}_{k} \text{vec}(G_k), \mathcal{M}_{k}^2 \text{vec}(G_k), \dots, \mathcal{M}_{k}^{J-1} \text{vec}(G_k)\}.
\end{equation*}
Seeking solutions in these subspaces ensures rapid error reduction and orthogonality in the iterates.
However, as $\dimbasis$ and $\lenparam$ increase, constructing $\mathcal{M}_{k}$ becomes computationally costly, especially for $k=2$. The matrix size grows rapidly, resulting in a large, dense operator that demands significant computational resources. To overcome this challenge, we solve the generalized Sylvester equation by applying the conjugate-gradient method directly to problem \eqref{MLROM Sylvester_problem}, hence avoiding the need to construct the full Kronecker product explicitly. This approach dramatically reduces the computational cost and memory requirements and makes the method scalable to large-dimensional applications.

\subsection{Conjugate-gradient algorithm to solve I-GILD problem}

When adapting the conjugate-gradient (CG) method to solve the generalized Sylvester equation \eqref{linear system Sylvester Pb}, we rely on iterative updates in the Krylov subspace. The primary goal is to solve a system of equations involving the generalized Sylvester structure, where the solution is iteratively updated along search directions that are mutually conjugate with respect to the Frobenius inner product. Let us detail the key aspects of this process. For the sake of simplicity, we omit the subscript $k$ in the subsequent description.

We begin with an initial guess for the solution, denoted by $X^{(0)},$ which can be initialized to zero or any other approximation. The residual $R^{(0)}$ is computed as the difference between the right-hand side $G$ and the action of the operator $\mathcal{S},$ such that  $\mathcal{S}(X) = \sum_{m=0}^{\lenparam} A_m X D_{m}$. That is
\begin{equation*}
    R^{(0)} = G - \mathcal{S}(X^{(0)}),
\end{equation*}
At each iteration $j,$ we generate a sequence of Krylov subspaces that expand the solution space in which the CG method searches for an optimal solution. Thus the $j$-th Krylov subspace is given by
\begin{equation*}
    \mathscr{K}_j(\mathcal{S}, R^{(0)}) = \text{span}\{R^{(0)}, \mathcal{S}(R^{(0)}), \mathcal{S}^2(R^{(0)}), \dots, \mathcal{S}^{j-1}(R^{(0)})\}.
\end{equation*}
with $\mathcal{S}^{j}$ as the operator defined recursively $\mathcal{S}\circ\mathcal{S}^{j-1}.$ This subspace contains the residual and its transformations under the operator $\mathcal{S},$ which allows us to express the solution as a linear combination of the basis vectors of the Krylov subspace. The key is to find the optimal coefficients for this combination by minimizing the residual Frobenius-norm.

To advance the CG process, we define a search direction $P^{(j)}$ along which the next step will be taken. The initial direction $P^{(0)}$ is taken equal to the initial residual $R^{(0)}.$ The search directions must be conjugate with respect to the operator $\mathcal{S}$, meaning that for any two distinct search directions $P^{(i)}$ and $P^{(l)}$ (with $i \neq l$), we must have
\begin{equation*}
    \langle P^{(i)}, \mathcal{S}(P^{(l)}) \rangle_F = 0,
\end{equation*}
where $\langle \cdot, \cdot \rangle_F$ denotes the Frobenius inner product defined for two matrices $ \bar{A}$ and $\bar{B}$ by $\langle \bar{A}, \bar{B} \rangle_F = \text{tr}(\bar{A}^T \bar{B}).$ This conjugacy with respect to the Frobenius inner product ensures that the search directions are mutually orthogonal in terms of reducing the residual of the Sylvester equation along the   directions $P^{(i)}.$

Now, with the direction $P^{(j)}$ established, a scalar $\alpha^{(j)}$ must be computed to determine the step size in this direction. This scalar is chosen such that the new residual is minimized in the direction of $P^{(j)}.$ The expression for $\alpha^{(j)}$ is given by
\begin{equation*}
    \alpha^{(j)} = \frac{||R^{(j)}||_F^2}{\langle P^{(j)}, \mathcal{S}(P^{(j)}) \rangle_F},
\end{equation*}
%
Once $\alpha^{(j)}$ is determined, the solution $ X^{(j+1)}$ is updated according to  
\begin{equation*}
    X^{(j+1)} \longleftarrow X^{(j)} + \alpha^{(j)} P^{(j)}.
\end{equation*}
This update takes a step in the direction of $P^{(j)},$ improving the approximate solution.
After the solution is updated, the residual is recalculated as
\begin{equation*}
    R^{(j+1)} \longleftarrow R^{(j)} - \alpha^{(j)} \mathcal{S}(P^{(j)}),
\end{equation*}
which updates the residual to reflect the current error after a step in the search direction.
To ensure that the next search direction $P^{(j+1)}$ remains conjugate to the previous ones, we introduce a scalar $\beta^{(j)},$ which is computed as
\begin{equation*}
    \beta^{(j)} = \frac{||R^{(j+1)}||_F^2}{||R^{(j)}||_F^2}.
\end{equation*}
This scalar ensures that the new search direction $P^{(j+1)}$ is a combination of the new residual  $R^{(j+1)}$ and the previous search direction $P^{(j)}:$
\begin{equation*}
    P^{(j+1)} \longleftarrow R^{(j+1)} + \beta^{(j)} P^{(j)}.
\end{equation*}
%
This iterative process continues until the residual norm $||R^{(j)}||_F$ falls below a predefined tolerance $\varepsilon,$ indicating satisfactory convergence. At this point, the solution $X^{(j)}$ approximates the true solution to the generalized Sylvester problem, and the Frobenius inner product guarantees that the search directions have remained orthogonal throughout the iterations.
%
%
The CG iterative process of the generalized Sylvester problem is summarized in algorithm \ref{alg:CG_Sylvester}. It is applied in two layers: to approximate (i) the linear model's coefficients $X_1,$ by solving $\widetilde{\mathcal{P}}_1$ and (ii) the quadratic coefficients $X_2$ by solving $\widetilde{\mathcal{P}}_2$.

\begin{algorithm}
    Initialize $ X^{(0)} $, for instance $ X^{(0)} = 0 $ \;
    $R^{(0)} = G - \mathcal{S}(X^{(0)})$ \;
    $P^{(0)} = R^{(0)}$ \;
    $j = 0$ \;
    
    \While{$|| R^{(j)} ||_F > \varepsilon$}{
    	Compute $ S^{(j)} = \mathcal{S}(P^{(j)})$\;
        Compute $ \alpha^{(j)} = {||R^{(j)}||_F^2} /\langle P^{(j)}, S^{(j)} \rangle_F $ \;
        $X^{(j+1)} = X^{(j)} + \alpha^{(j)} P^{(j)}$ \;
        $R^{(j+1)} = R^{(j)} - \alpha^{(j)} S^{(j)}$ \;

        \If{$||R^{(j+1)}||_F < \varepsilon$}{
            \textbf{break}
        }
        Compute $ \beta^{(j)} = {||R^{(j+1)}||_F^2} / {||R^{(j)}||_F^2} $ \;
        $P^{(j+1)} = R^{(j+1)} + \beta^{(j)} P^{(j)}$ \;
        $j \longleftarrow j + 1$ \;
    }
    \caption{Conjugate-gradient solver for the generalized Sylvester problem}
\label{alg:CG_Sylvester}
\end{algorithm}


\subsection{Convergence analysis of the I-GILD problem}

In this section, we analyze the convergence behavior of the conjugate-gradient (CG) method when applied to the generalized Sylvester equation \eqref{MLROM Sylvester_problem}. In this problem,  $A_0$ and $D_{k,0}$ are the diagonal regularization matrices, and $A_m$ and $D_{k,m}$ for $m \neq 0$ are symmetric and positive semi-definite (SPSD). 
The rate of convergence of the CG method depends on the condition number $\kappa(\mathcal{M}_k)$ of the matrix $\mathcal{M}_k$ given by \eqref{Mk_kron}, which is influenced by the eigenvalues of the matrices $A_m$ and $D_{k,m}.$ 
Since $A_m$ and $D_{k,m}$ are SPSD matrices, they have a subset of eigenvalues that are exactly zero. These zero eigenvalues correspond to spatial directions along which the matrix has no action (i.e., a null-space). In such cases, the matrix is not invertible in those directions, and any attempt to compute the condition number using a zero eigenvalue would result in an undefined (or infinite) value.
However, in practice, the CG method operates on the subspace where the matrices $A_m$ and $D_{k,m}$ are positive definite, i.e., on the subspace where the eigenvalues are strictly positive. This is a consequence of the CG method progressing, by design, through the Krylov subspaces that are spanned by spatial directions that correspond to non-zero eigenvalues. Directions corresponding to zero eigenvalues do not affect the convergence of the method since they do not contribute any change to the solution. The convergence is governed by the non-zero eigenvalues, which represent the effective conditioning of the problem.
Hence, we define the condition number as the ratio of the largest to the smallest, non-zero eigenvalue of the system \cite{trefethen1997numerical}. Specifically, we express the condition number as:

\begin{equation*}
\kappa(\mathcal{M}_k) = \frac{\max_m \left\{ \lambda_{\max}(A_m) \lambda_{\max}(D_{k,m}) \right\}}{\min_{m} \left\{ \lambda_{\min}^+(A_m) \lambda_{\min}^+(D_{k,m}) \right\}},
\end{equation*}
where $\lambda_{\min}^+(A_{k,m})$ and $\lambda_{\min}^+(D_{k,m})$ denote the smallest non-zero eigenvalues of $A_m$ and $D_{k,m}.$ The asymptotic convergence of the CG method is closely tied to the condition number $\kappa(\mathcal{M}_k).$ 
Note that since $\basis{m}{}{}$ are orthogonal, the nonzero eigenvalues of $A_m$ are all equal to one, simplifying the condition number of $M_k$ to  
\begin{equation*}
\kappa(\mathcal{M}_k) = \frac{\max_m \left\{ \lambda_{\max}(D_{k,m}) \right\}}{\min_{m} \left\{ \lambda_{\min}^+(D_{k,m}) \right\}},
\end{equation*}
This expression highlights the fact that the conditioning for the I-GILD problem is directly tied to the non-zero eigenvalues distribution of the matrices $D_{k,m}$.
It's well known that after $j$ iterations of the CG method, the convergence error $e_X^{(j)} = X^{(j)} - X^*,$ where $X^*$ stands for the exact solution, is bounded by
\begin{equation*}
    \frac{|| e_{X_k}^{(j)} ||_F}{|| e_{X_k}^{(0)} ||_F} \leq 2 \left( \frac{\sqrt{\kappa(\mathcal{M}_k)} - 1}{\sqrt{\kappa(\mathcal{M}_k)} + 1} \right)^j.
\end{equation*}
This error bound underscores the fact that the convergence rate improves significantly when the condition number is reduced. A lower condition number ensures faster reduction of the error, making the method more efficient. 

\subsection{Computational complexity}

In the following section, we discuss the computational complexity of solving the Sylvester problem for the quadratic coefficients, $k = 2,$ as this case is the most computationally expensive step in terms of matrix-matrix multiplications during each iteration of the CG method. We recall that the matrix $A_m$ is of size $\lenbasis \times \lenbasis,$ while $D_{2,m}$ is of size $\lenbasis^2 \times \lenbasis^2,$ and the search direction matrix $P^{(j)}$ is of size $\lenbasis \times \lenbasis^2.$

In each iteration, the product $A_m P^{(j)}$ is first computed, which has a computational complexity of $O(\lenbasis^4),$ since $P^{(j)}$ is of size $\lenbasis \times \lenbasis^2$ and $A_m$ is of size $\lenbasis \times \lenbasis.$ After that, the resulting matrix is multiplied by $D_{k,m},$ which has a complexity of $O(\lenbasis^5),$ given the size of $D_{2,m}$ is $\lenbasis^2 \times \lenbasis^2.$
Since we sum over $m = 1, \dots, \lenparam,$ the total computational complexity per iteration comes to $O(\lenparam \lenbasis^5.$ 
The total computational complexity of the CG method also depends on the number $J$ of iterations required for convergence.
Therefore, the total complexity of the CG method is
\begin{equation*}
    O(J \cdot \lenparam^6 \cdot \dimbasis^5),
\end{equation*}
Besides the high cost of forming the matrix $D_{m,2},$ the conjugate-gradient iterations involving the computation of the operator $\mathcal{S}(P^{(j)})$ are particularly expensive, as they scale with $\lenbasis^5.$ This further contributes to a significant computational burden for large-scale problems.
However, in I-GILD context, it is possible to reduce the computational complexity in  by taking advantage of the structure of $D_{2,m},$ expressed through the Kronecker product. Note that the product $A_m P^{(j)} D_{2,m}$ can be expressed as $\basis{m}{}{} \basis{m}{}{}^T P^{(j)} \Psi_{2,m} \Psi_{2,m}^T$. So, instead of explicitly forming $D_{2,m} = \Psi_{2,m} \Psi_{2,m}^T,$ we can use the Kronecker product decomposition of $\Psi_{2,m}$ into three sparse matrices given by
\begin{equation}\label{Kron_decomposition_sparse}
    \Psi_{2,m} = \left( I_{\lenbasis} \otimes \basis{m}{}{} \right) 
    \left( \basis{m}{}{} V_m^T \otimes I_{\dimbasis} \right)
    \left( \sum_{n=0}^{\lensnap-1} \delta_n \otimes \tempcoef{m}{}{n} \right).
\end{equation}
We begin by multiplying $P^{(j)}$ from the left by $\basis{m}{}{}^T,$ where $\basis{m}{}{}$ has dimensions $\lenbasis\times \dimbasis.$ After this operation, the computational complexity is $O(\lenparam^3 \cdot \dimbasis^4).$ Next, we multiply the result by $\basis{m}{}{},$ yielding the same computational complexity.
Now, the matrix is ready for multiplication by $\Psi_{2,m}$ from the right. 
We first multiply by $\left( I_{\lenbasis} \otimes \basis{m}{}{} \right).$ The size of the matrix becomes $\lenbasis \times \lenbasis \cdot \dimbasis ,$ and the complexity of this operation is $ O(\lenparam^3 \cdot \dimbasis^4).$
Next, we multiply by $\left( \basis{m}{}{} V_m^T \otimes I_{\dimbasis} \right),$ where $V_m$ has dimensions $\lensnap \times \dimbasis.$ The computational complexity of this step is $ O(\lenparam^2 \cdot \dimbasis^3 \cdot \lensnap).$
Finally, we multiply by $\sum_{n=0}^{\lensnap-1} \delta_n \otimes \tempcoef{m}{}{n}.$ The matrix size changes to $\lenbasis \times \lensnap,$ and the computational complexity of this operation is  $O(\lensnap \cdot \dimbasis^3 \cdot \lenparam).$
After this, we proceed to multiply the result by $\Psi_{2,m}^T$ from the right. The transpose operation follows the same steps as for $\Psi_{2,m},$ thus having the same computational complexities as the forward multiplication.
Thus, the total complexity per iteration for executing the multiplication $A_m \mathcal{S}(P^{(j)}) D_{m,2}$ is 

\begin{equation*}
O(\lenparam^3 \cdot \dimbasis^4)
\quad + \quad
O(\lenparam^3 \cdot \dimbasis^3 \cdot \lensnap)
\quad + \quad 
O(\lenparam \cdot \dimbasis^3 \cdot \lensnap).
\end{equation*}
By accounting only for the dominant term in the evaluation of the cost, and considering that  $\dimbasis\leq\lensnap$, the total complexity of the CG iteration using the Kronecker decomposition \eqref{Kron_decomposition_sparse} is estimated as
\begin{equation*}
O(J \cdot \lenparam^3 \cdot \lensnap^4).
\end{equation*}
The above steps enable a reduction in computational complexity compared to explicitly forming $D_{2,m},$ which would have resulted in a complexity estimated as $O(J \cdot \lenparam^6 \cdot \lensnap^5).$ While not a drastic reduction, the above approach is nonetheless advantageous, as it avoids the need to explicitly form and store the large dense matrix $D_{2,m},$ which would otherwise be computationally prohibitive, especially in terms of memory requirements. The method allows for more efficient handling of large-scale problems while still managing the significant computational burden associated with the growth in $\dimbasis$, $\lenparam,$ and $J$.

\section{Error growth analysis}

In this section, we aim to derive an error bound for the assimilated model by expressing the error at time step $n,$ labeled $\| e_m^n \|,$ in terms of the initial error $\| e_m^0 \|$ and the residuals history $\bm{r}_m^t$, for $t\leq n$. This analysis will help us understand the error growth in time. In doing so, we use the fact that the operators of the learned model are Lipschitz continuous, allowing us to control the growth as discussed in the standard reduced-order-modeling literature \cite{quarteroni2015reduced, hesthaven2016certified, benner2015survey}.

The error at time step $n,$ denoted by $e_m^n,$ is the difference between the true POD $\tempcoef{m}{}{n}$ and predicted latent vectors $\tempcoefGILD{m}{}{n}$ given by
\begin{equation*}
    e_m^n = \tempcoef{m}{}{n} - \tempcoefGILD{m}{}{n}.
\end{equation*}
The evolution of the predicted latent variables is governed by the equation
\begin{equation*}
    \tempcoefGILD{m}{}{n} = \basis{m}{}{}^T \matLconcat \basis{m}{}{} \tempcoefGILD{m}{}{n-1} + \basis{m}{}{}^T \matQconcat B_{*m}^{n-1} \tempcoefGILD{m}{}{n-1}.
\end{equation*}
Substituting this expression into the error equation produces:
\begin{equation*}
    e_m^{n} = \tempcoef{m}{}{n} - \left( \basis{m}{}{}^T \matLconcat \basis{m}{}{} \tempcoefGILD{m}{}{n-1} + \basis{m}{}{}^T \matQconcat B_{*m}^{n-1} \tempcoefGILD{m}{}{n-1} \right).
\end{equation*}
Next, we introduce and subtract the corresponding high-fidelity terms to express the error evolution as
\begin{equation*}
    e_m^n = \bm{r}_m^{n} + \basis{m}{}{}^T \matLconcat \basis{m}{}{} e_m^{n-1} + \basis{m}{}{}^T \matQconcat \left( B_{m}^{n-1} \tempcoef{m}{}{n-1} - B_{*m}^{n-1} \tempcoefGILD{m}{}{n-1} \right),
\end{equation*}
where $\bm{r}_m^{n}$ is the learning residual at time step $n$
\begin{equation*}
    \bm{r}_m^{n} = \tempcoef{m}{}{n} - \left( \basis{m}{}{}^T \matLconcat \basis{m}{}{} \tempcoef{m}{}{n-1} + \basis{m}{}{}^T \matQconcat B_m^{n-1} \tempcoef{m}{}{n-1} \right).
\end{equation*}
To further simplify the error expression, we need to simplify the quadratic difference $B_{m}^{n-1} \tempcoef{m}{}{n-1} - B_{*m}^{n-1} \tempcoefGILD{m}{}{n-1}$. For this, recall the structure of the matrices $B_m^n$ and $B_{*m}^n$
\begin{equation*}
    B_{m}^{n-1} = \left( \basis{m}{}{} \tempcoef{m}{}{n-1} \right) \otimes \basis{m}{}{}, \qquad B_{*m}^{n-1} = \left( \basis{m}{}{} \tempcoefGILD{m}{}{n-1} \right) \otimes \basis{m}{}{}.
\end{equation*}
Substituting these expressions into the quadratic difference above yields
\begin{equation*}
    B_{m}^{n-1} \tempcoef{m}{}{n-1} - B_{*m}^{n-1} \tempcoefGILD{m}{}{n-1} = \left( \basis{m}{}{} e_m^{n-1} \right) \otimes \basis{m}{}{} \tempcoef{m}{}{n-1} + \left( \basis{m}{}{} \tempcoef{m}{}{n-1} \right) \otimes \basis{m}{}{} e_m^{n-1} - \left( \basis{m}{}{} e_m^{n-1} \right) \otimes \basis{m}{}{} e_m^{n-1}.
\end{equation*}
Using this expansion in the error expression, we can derive an error bound that quantifies the error growth of the identified quadratic model at step $t_n$ in terms of the previous residuals and errors. Thus, we use the fact that $X_1$ and $X_2,$ as linear operators, naturally satisfy the Lipschitz continuity property. Specifically, the Lipschitz constants are given by $L_{X_1} = \sigma_{\max}(\matLconcat)$ and $L_{X_2} = \sigma_{\max}(X_2),$ respectively. Moreover, since $\basis{m}{}{}$ has orthonormal columns, it satisfies $\basis{m}{}{}^T \basis{m}{}{} = I_{\dimbasis}.$ This key property ensures that multiplication by $\basis{m}{}{}$ preserves the $2$-norm for any vector $x \in \mathbb{R}^{\dimbasis},$ i.e., $\| \basis{m}{}{} x \|_2 = \| x \|_2.$ Furthermore, multiplying by $\basis{m}{}{}^T,$ which maps vectors from $\mathbb{R}^{\lenbasis}$ to $\mathbb{R}^{\dimbasis},$ contracts or preserves the norm: $\| \basis{m}{}{}^T x \|_2 \leq \| x \|_2.$ Thus, for the linear term, we conclude
\begin{equation*}
    \| \basis{m}{}{}^T \matLconcat \basis{m}{}{} e_m^{n-1} \| \leq L_{X_1} \| e_m^{n-1} \|.
\end{equation*}
Bouncing back to the quadratic terms, we use $\| \tempcoef{m}{}{n-1} \| \leq 1$ and the Lipschitz condition on $X_2$ to obtain the bound
\begin{equation*}
    \| \basis{m}{}{}^T \matQconcat \left( \basis{m}{}{} e_m^{n-1} \otimes \basis{m}{}{} \tempcoef{m}{}{n-1} \right) \| \leq L_{X_2} \| e_m^{n-1} \|,
\end{equation*}
\begin{equation*}
    \| \basis{m}{}{}^T \matQconcat \left( \basis{m}{}{} \tempcoef{m}{}{n-1} \otimes \basis{m}{}{} e_m^{n-1} \right) \| \leq L_{X_2} \| e_m^{n-1} \|,
\end{equation*}
\begin{equation*}
    \| \basis{m}{}{}^T \matQconcat \left( \basis{m}{}{} e_m^{n-1} \otimes \basis{m}{}{} e_m^{n-1} \right) \| \leq L_{X_2} \| e_m^{n-1} \|^2.
\end{equation*}
Finally, the total error bound can be written as
\begin{equation*}
    \| e_m^n \| \leq \| \bm{r}_m^{n} \| + L_{X_1} \| e_m^{n-1} \| + 2 L_{X_2} \| e_m^{n-1} \| + L_{X_2} \| e_m^{n-1} \|^2.
\end{equation*}
To achieve a recursive form of the error, we need to handle the quadratic term $\| e_m^{n-1} \|^2,$ in the right hand side member of the bound. To this end, we introduce the assumption $\| e_m^{n-1} \| \ll 1,$ common in reduced-order models for small initial errors and short-term evolution. This yields $\| e_m^{n-1} \|^2 \leq \| e_m^{n-1} \|$, and thus simplifies the error bound to
\begin{equation*}
    \| e_m^n \| \leq \| \bm{r}_m^{n} \| + \xi \| e_m^{n-1} \|,
\end{equation*}
where $\xi = L_{X_1} + 3 L_{X_2}.$ Expanding the error recursively gives:
\begin{equation*}
    \| e_m^n \| \leq \sum_{t=0}^{n-1} \xi^t \| \bm{r}_m^{n-t} \| + \xi^n \| e_m^0 \|.
\end{equation*}
This bound indicates that the error at time step $n$ is inherently influenced by both the initial error and the residuals from previous steps. The parameter $\xi$ plays a crucial role in controlling the error growth over time. Provided that the initial errors are small, when $\xi < 1$, the error remains bounded and stable. Conversely, if $\xi > 1$, the error may grow exponentially. Nevertheless, even in scenarios where $\xi > 1$, the model can still perform effectively, provided that the residuals are kept sufficiently small.

\section{Numerical Results}
{
In this section, we present numerical tests to assess the performance of the proposed I-GILD approach on two flow cases. The first test case examines the flow around an Ahmed body, where the varying parameter is the rear slant angle, with data generated from numerical simulations. The second, more challenging test case involves flow in a cylindrical lid-driven cavity at different Reynolds numbers, using experimental data obtained from PIV measurements. The inclusion of experimental data in the second test case introduces real-world complexities, such as measurement noise and inconsistencies, providing a more rigorous test of the I-GILD model’s robustness and its ability to generalize beyond synthetic datasets.
To ensure meaningful error comparisons between predictions and the high-fidelity data, we apply a time-shifting calibration through an orthogonal transformation to account for potential phase shifts and structural misalignment in the flow fields. This calibration process addresses common challenges in flow data, such as temporal shifts in vortex structures or rotational misalignments, by aligning the dominant structures in the predicted data matrices to the high-fidelity data. Specifically, this calibration is obtained by solving the optimization problem 

\begin{equation}
    \min_{Q \in \mathcal{O}_{\lensnap+1}} \, \|\widetilde{\mathcal{U}} Q - \mathcal{U}\|_F^2,
\end{equation}
where $\widetilde{\mathcal{U}}$ denotes the predicted flow field matrix produced by the I-GILD model, that needs calibration, $\mathcal{U}$ represents the high-fidelity matrix, and $Q$ is an orthogonal matrix within the set $\mathcal{O}_{\lensnap+1}$. {Such an approach ensures that the error measure reflects structural similarity in the flow fields rather than focusing on minor temporal misalignment, making it robust to phase and rotational shifts that are often present in the prediction of complex flow dynamics.
}

}

\subsection{Ahmed body flow with variable rear slant angle}

This study explores the ability of the I-GILD method to predict the 2D flow around an Ahmed body \cite{AhmedBody1984}, with varying geometric configurations, specifically focusing on changes in the slant angle $ \alpha $ of the upper rear section. The Ahmed body has a length of $ \ell_2 = 1.044 \, \text{m} $ and a width of $ h_1 = 0.228 \, \text{m} $, positioned $ \ell_1 = 1.05 \, \text{m} $ from the inlet and $ \ell_3 = 2.45 \, \text{m} $ from the outlet. The distance between the body and the road is $ 0.05 \, \text{m} $, while the gap between the upper boundary and the top wall is $ 1.172 \, \text{m} $. A constant horizontal inlet velocity of $ U = 14.7 \, \text{m/s} $ drives the flow, with no-slip boundary conditions applied at the road and body surfaces, and symmetry and zero gradient conditions imposed at the upper wall and outlet, respectively.
The Reynolds number $ Re $ is computed based on the Ahmed body length $ \ell_2 $ as $ Re = U \ell_2 / \nu $, where $ \nu = 1.47 \times 10^{-5} \, \text{m}^2/\text{s} $ is the kinematic viscosity of air at $ 15^{\circ}C $. This results in a Reynolds number of $ Re = 1,044,000 $.
For the simulations, three slant angles were used for training: $ 7.998^{\circ} $, $ 11.848^{\circ} $, and $ 15.556^{\circ} $. The test cases employed slant angles of $ 9.944^{\circ} $ and $ 14.085^{\circ} $. These simulations were conducted using OpenFOAM with the SpalartAllmarasIDDES turbulence model on a mesh containing 14,450 nodes, with an adaptive time step initialized to $ \Delta t = 0.001 \, \text{s} $.

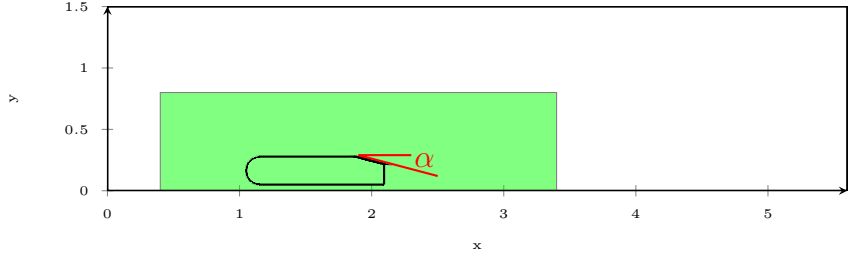
\begin{figure}
\centering
\begin{subfigure}{0.9\textwidth}
\begin{tikzpicture}
\Large
  \begin{axis}[
    xmin = 0., xmax = 5.6, ymin = 0, ymax = 1.5,
    axis lines = left,
    width=0.9\textwidth, 
    height=0.32\textwidth, 
    style={font=\tiny},
    xlabel={x},
    ylabel={y},
  ]
    \addplot[black, opacity=0.5, fill=green, fill opacity=0.5] coordinates {(0.4, 0) (3.4, 0) (3.4, 0.8) (0.4, 0.8) (0.4, 0)};
    
    \addplot[black, smooth, style={thick, font=\tiny}] table{./ahmed_body_points_ordered.txt};
    
    \draw[black, thick] (axis cs:0,0) rectangle (axis cs:5.6,1.5);
    
    \coordinate (rear_corner) at (axis cs:1.9, 0.29);  
    \coordinate (slant_end) at (axis cs:2.5, 0.12);   

    \draw[red, thick] (rear_corner) -- (slant_end);  
    \draw[red, thick] (rear_corner) -- (axis cs:2.3, 0.29);  

    \node[red] at (axis cs:2.4, 0.25) {\large $\alpha$};

  \end{axis}
\end{tikzpicture}
\end{subfigure}
\caption{Configuration of Ahmed-body flow. The parameter is varied through the rear slant angle $\theta$. The red shaded area represents the region of interest within the domain where data are collected for use in GILD method. The dimensions of the frame and its position relative to the Ahmed body are shown to be within the squared subdomain $[0.4,3.4]\times[0,0.8]$.}
\label{Fig. geometry ABR}
\end{figure}

To handle geometric variations, specifically changes in the rear slant, an immersed boundary approach was used, as detailed in \cite{GILD2024}. This technique projects the velocity fields onto a reference rectangular domain refined near the body, with the fluid assumed to occupy the entire body’s interior with zero velocity.
In constructing the GILD model, a subdomain around the Ahmed body, illustrated in Figure \ref{Fig. geometry ABR}, was used. Velocity fields were projected onto this subdomain to generate POD bases. A total of $100$ velocity snapshots, taken between  $4.5 \, \text{s}$  and  $12 \, \text{s}$  with a time increment of  $\Delta t = 0.075 \, \text{s}$, were used. Of these, $70$ were allocated for training, and the remaining $30$ for testing the models in forecast mode. This setup provides a robust evaluation of both GILD and I-GILD, particularly in assessing their ability to extrapolate dynamics beyond the training data. For the latent space, a POD truncated order of $35$ was used, ensuring that $99.9\%$ of the cumulative POD energy was captured.

\begin{figure}[hbtp!]
    \centering
    \begin{subfigure}{0.45\textwidth}
        \includegraphics[width=\textwidth]{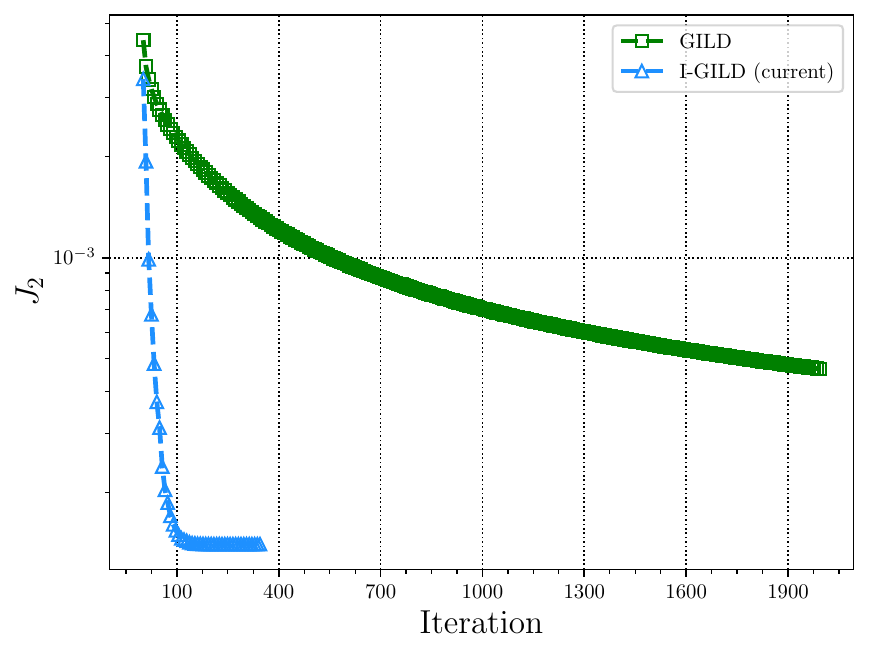}
        \caption{Functional decay}
    \end{subfigure}%
    \begin{subfigure}{0.45\textwidth}
        \includegraphics[width=\textwidth]{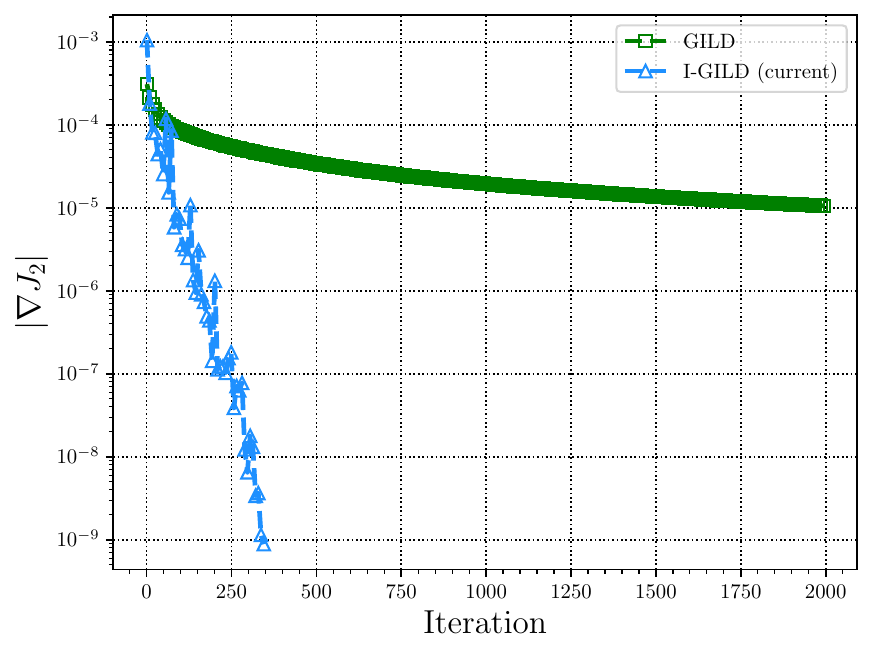}
        \caption{Residual decay}
    \end{subfigure}

    \caption{Convergence comparison between GILD and I-GILD for learning the quadratic coefficients in the model, demonstrating the evolution of functional $J_2$ and the magnitude of its gradient $\nabla J_2$ over iterations.}

    \label{Fig. convergence GILD comparison ABR}
\end{figure}

The performance of I-GILD, based on the Conjugate Gradient (CG) method, is compared to the original GILD model. As shown in Figure \ref{Fig. convergence GILD comparison ABR}, I-GILD exhibits significantly faster convergence during training. It achieves an order of magnitude reduction in the objective functional after only 100 iterations, outperforming the original GILD, which uses a steepest descent optimization method. This improvement is a direct result of the more efficient error minimization achieved by the CG method, validating its theoretical advantages over steepest descent for this problem.

\begin{figure}[hbtp!]
\centering
\begin{subfigure}{0.32\textwidth}
    \includegraphics[width=\textwidth]{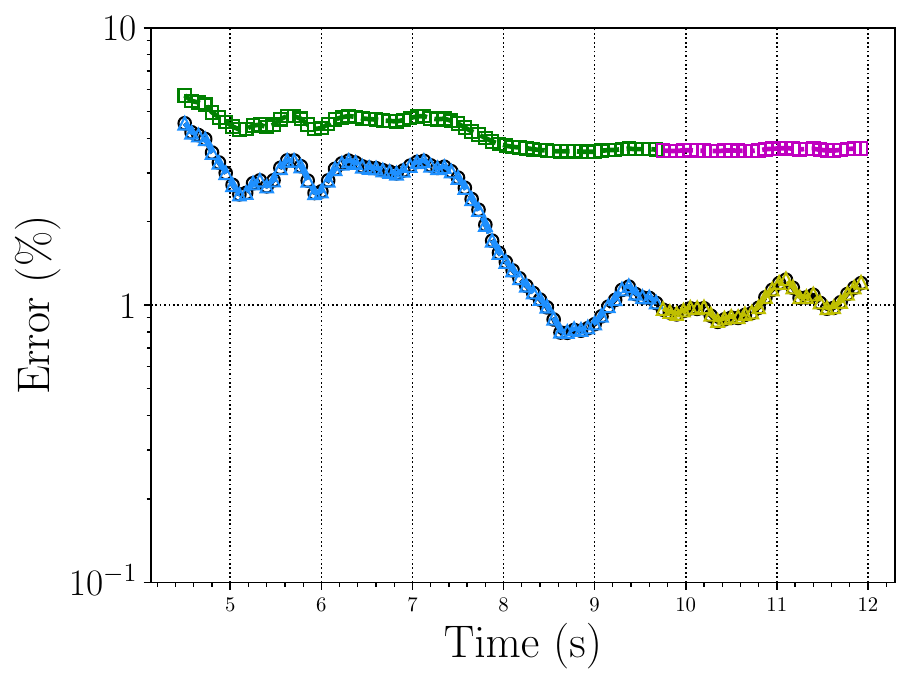}
    \caption{training angle $7.998^{\circ}$}
\end{subfigure}%
\begin{subfigure}{0.32\textwidth}
    \includegraphics[width=\textwidth]{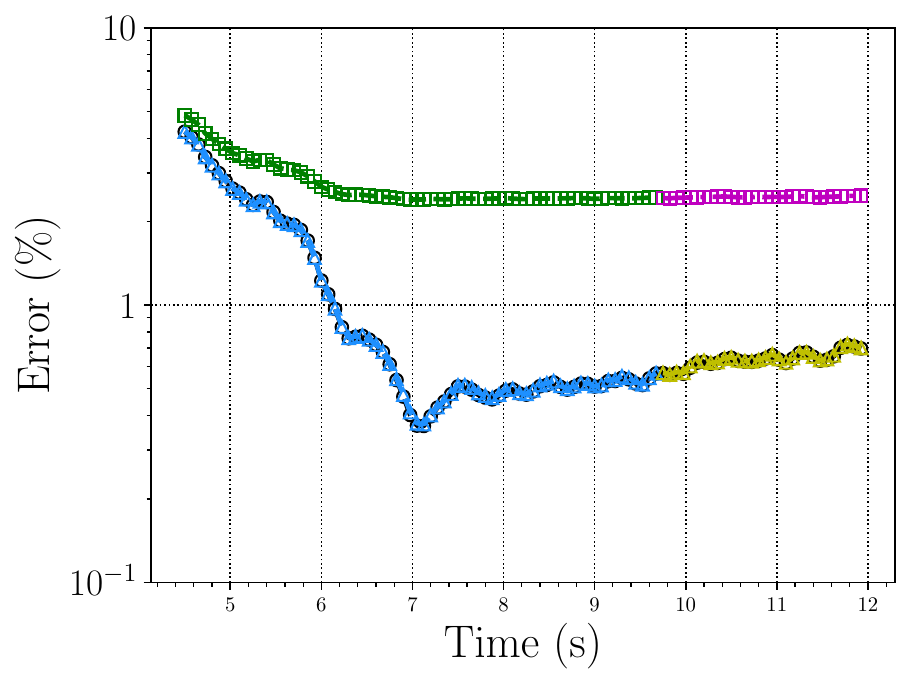}
    \caption{training angle $11.848^{\circ}$}
\end{subfigure}%
\begin{subfigure}{0.32\textwidth}
    \includegraphics[width=\textwidth]{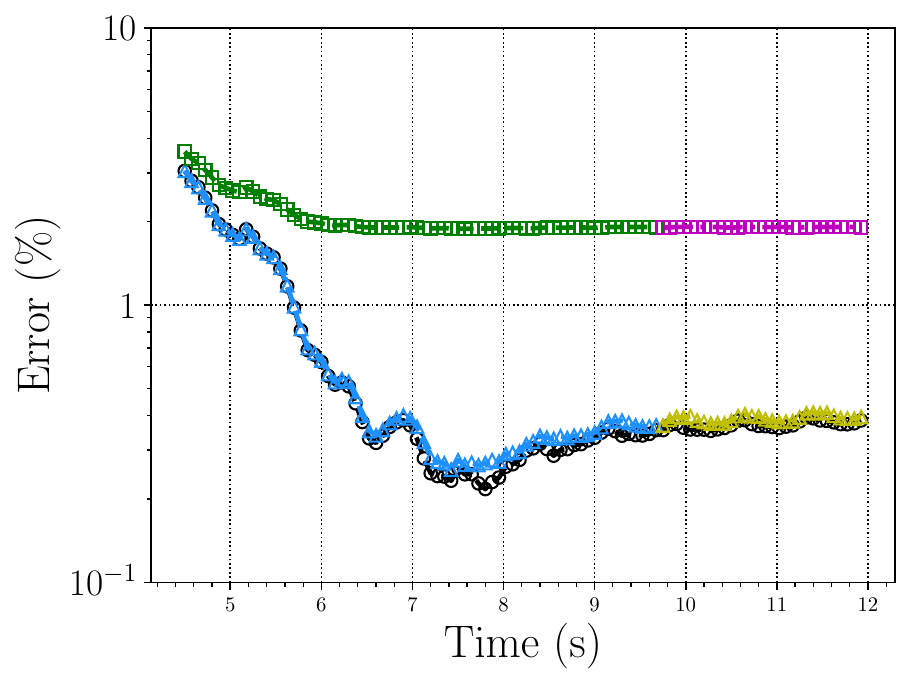}
    \caption{training angle $15.556^{\circ}$}
\end{subfigure}

\begin{subfigure}{0.32\textwidth}
    \includegraphics[width=\textwidth]{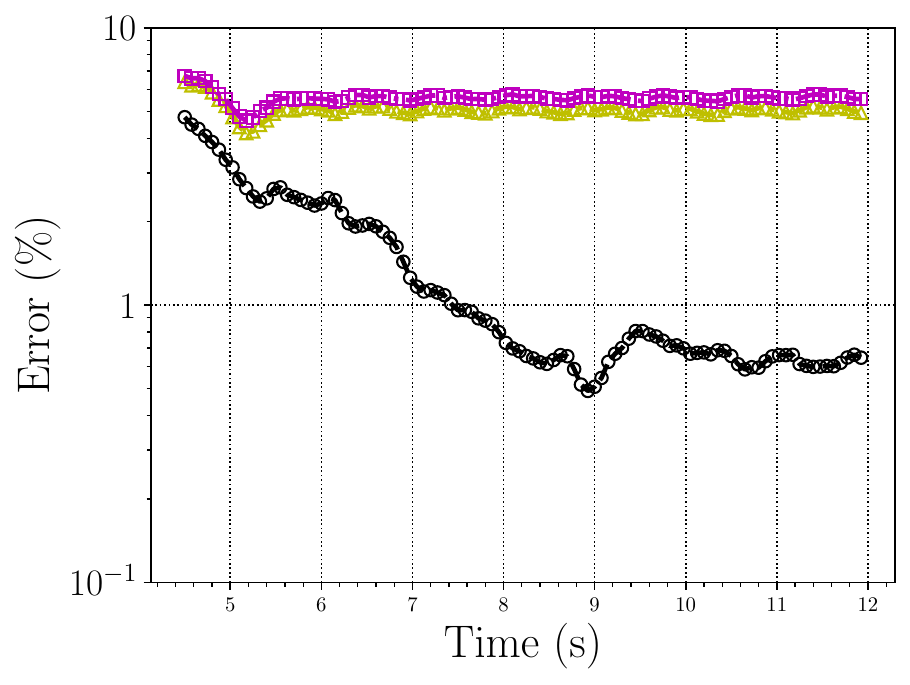}
    \caption{test angle $9.944^{\circ}$}
\end{subfigure}%
\begin{subfigure}{0.32\textwidth}
    \includegraphics[width=\textwidth]{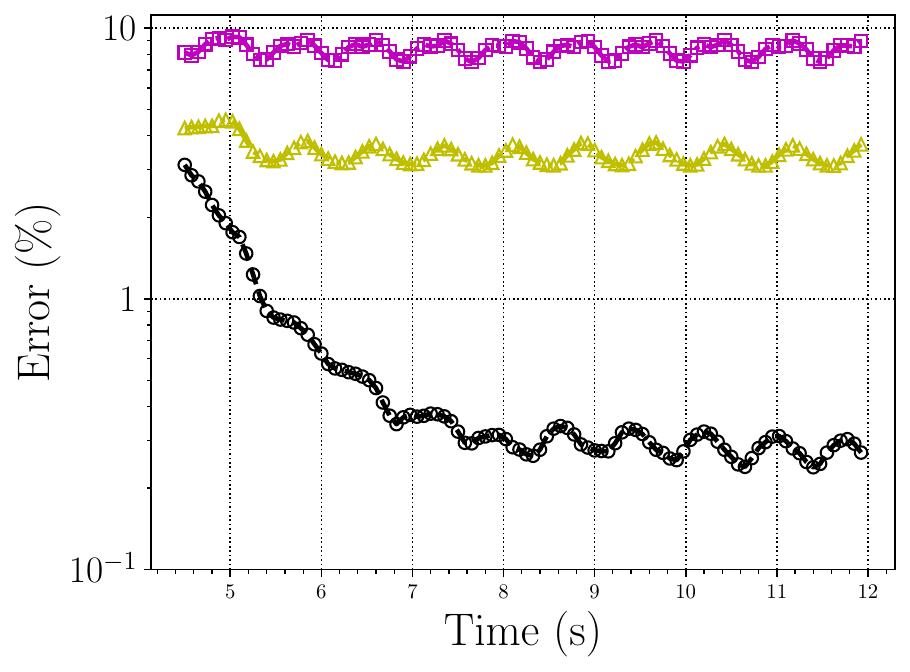}
    \caption{test angle $14.085^{\circ}$}
\end{subfigure}

\includegraphics[width=\textwidth]{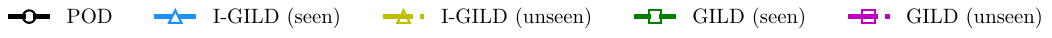}

\caption{Percentage of error with respect to time for the training and test slant angles.}
\label{Fig. Errors wrt time Ahmed Body}
\end{figure}

\begin{figure}[hbtp!]
\centering
\begin{subfigure}[b]{\textwidth}
    \centering
    \includegraphics[width=0.23\linewidth]{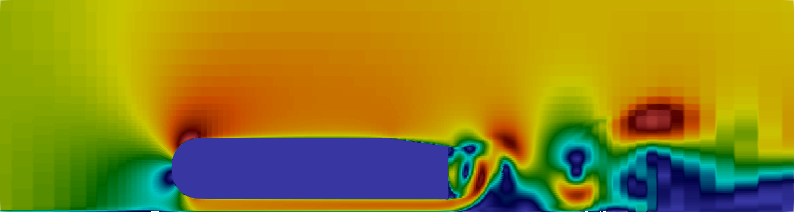}%
    \hspace*{0.01\linewidth}
    \includegraphics[width=0.23\linewidth]{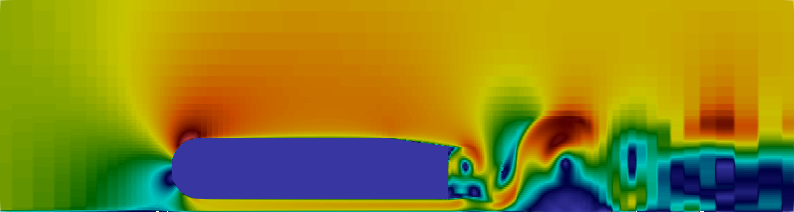}%
    \hspace*{0.01\linewidth}
    \includegraphics[width=0.23\linewidth]{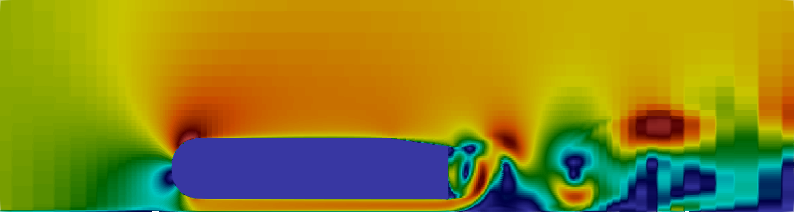}%
    \hspace*{0.01\linewidth}
    \includegraphics[width=0.23\linewidth]{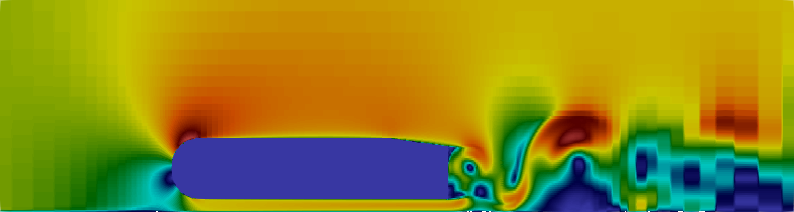}
    \caption{High fidelity training flow : slant angle $\alpha = 7.998^{\circ}$}
\end{subfigure}
\begin{subfigure}[b]{\textwidth}
    \centering
    \includegraphics[width=0.23\linewidth]{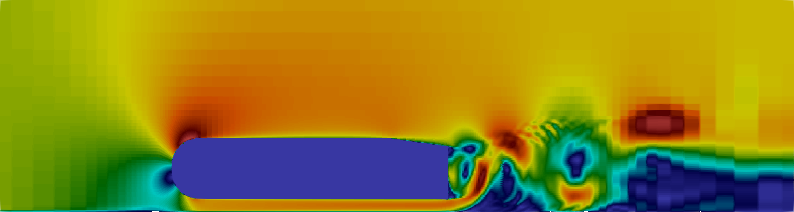}%
    \hspace*{0.01\linewidth}
    \includegraphics[width=0.23\linewidth]{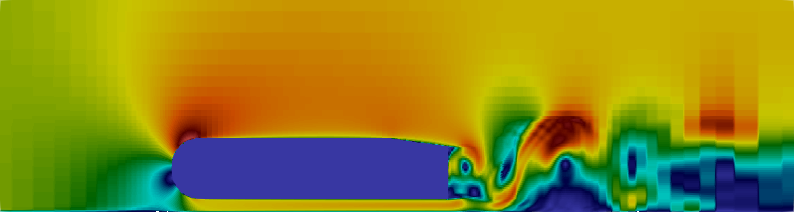}%
    \hspace*{0.01\linewidth}
    \includegraphics[width=0.23\linewidth]{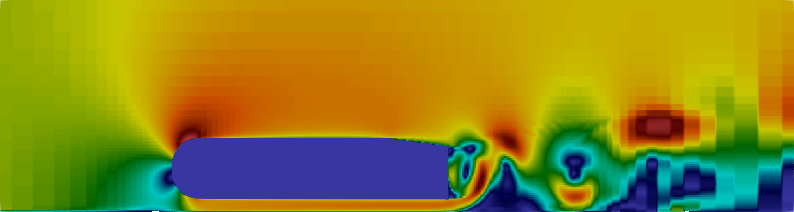}%
    \hspace*{0.01\linewidth}
    \includegraphics[width=0.23\linewidth]{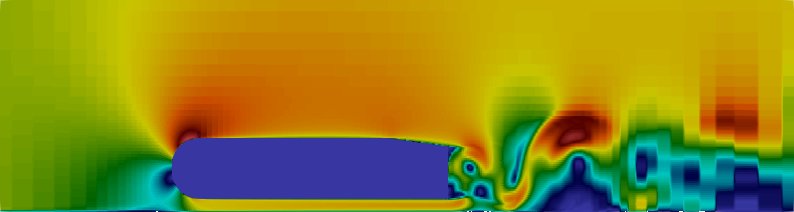}
    \caption{I-GILD}
\end{subfigure}

\caption{Comparison of the I-GILD and POD reconstructed flows alongside the high-fidelity flow for the training slant angle $\alpha = 7.998^{\circ}$, at time instants of $t=4.5$ s, $t=6.75$ s, $t=9$ s, and $t=11.25$ s, respectively from left to right.}
\label{Fig. AB flow test 7.998}
\end{figure}

\begin{figure}[hbtp!]
\centering
\begin{subfigure}[b]{\textwidth}
    \centering
    \includegraphics[width=0.23\linewidth]{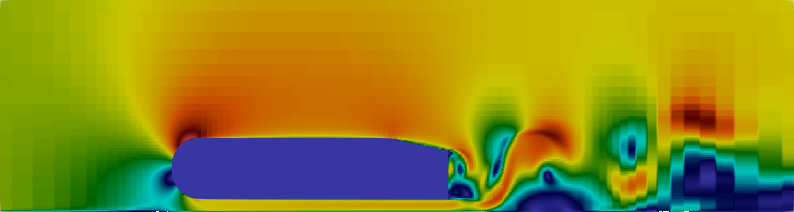}%
    \hspace*{0.01\linewidth}
    \includegraphics[width=0.23\linewidth]{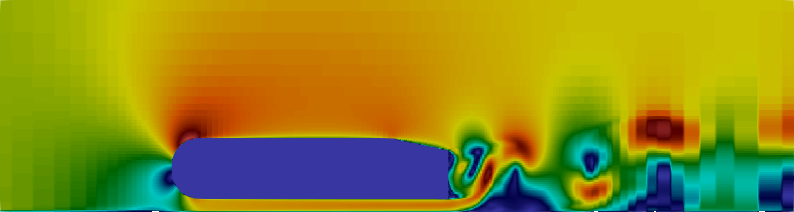}%
    \hspace*{0.01\linewidth}
    \includegraphics[width=0.23\linewidth]{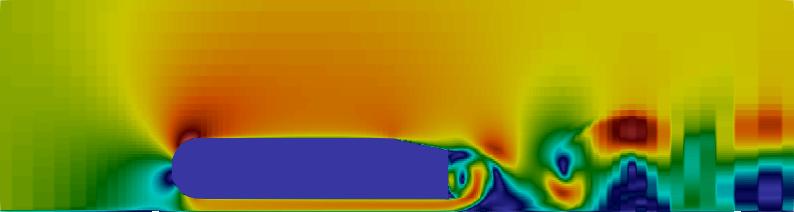}%
    \hspace*{0.01\linewidth}
    \includegraphics[width=0.23\linewidth]{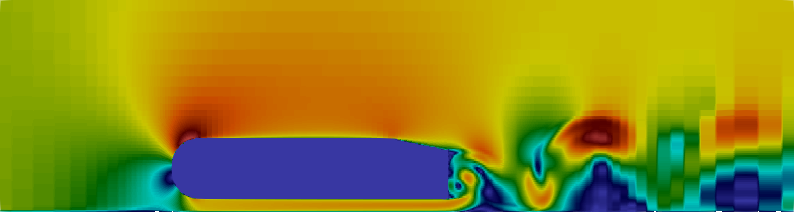}
    \caption{High fidelity training flow : slant angle $\alpha = 11.848^{\circ}$}
\end{subfigure}
\begin{subfigure}[b]{\textwidth}
    \centering
    \includegraphics[width=0.23\linewidth]{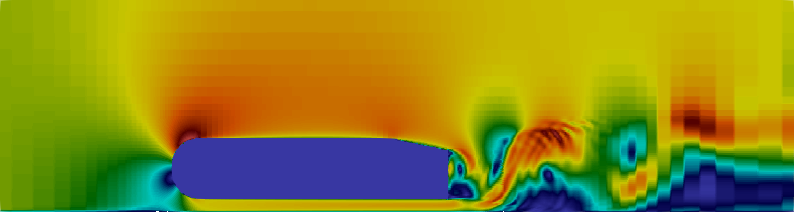}%
    \hspace*{0.01\linewidth}
    \includegraphics[width=0.23\linewidth]{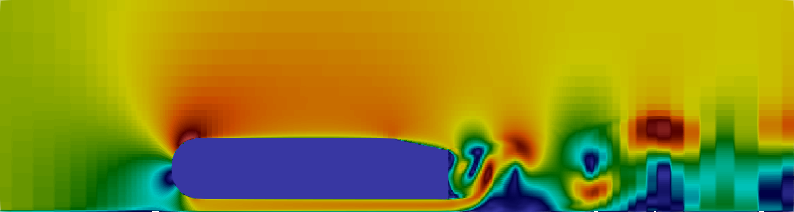}%
    \hspace*{0.01\linewidth}
    \includegraphics[width=0.23\linewidth]{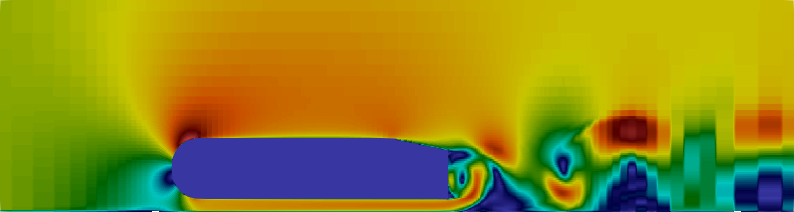}%
    \hspace*{0.01\linewidth}
    \includegraphics[width=0.23\linewidth]{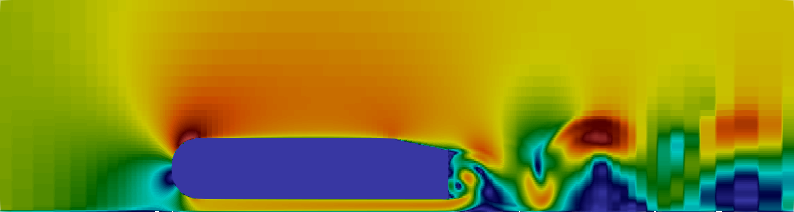}
    \caption{I-GILD}
\end{subfigure}

\caption{Comparison of the I-GILD and POD reconstructed flows alongside the high-fidelity flow for the training slant angle $\alpha = 11.848^{\circ}$, at time instants of $t=4.5$ s, $t=6.75$ s, $t=9$ s, and $t=11.25$ s, respectively from left to right.}
\label{Fig. AB flow test 11.848}
\end{figure}

\begin{figure}[hbtp!]
\centering
\begin{subfigure}[b]{\textwidth}
    \centering
    \includegraphics[width=0.23\linewidth]{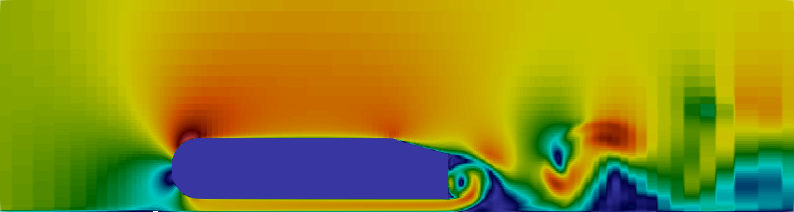}%
    \hspace*{0.01\linewidth}
    \includegraphics[width=0.23\linewidth]{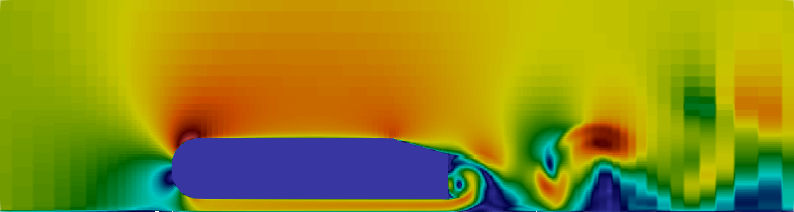}%
    \hspace*{0.01\linewidth}
    \includegraphics[width=0.23\linewidth]{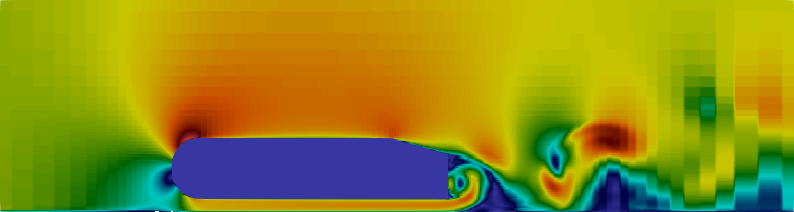}%
    \hspace*{0.01\linewidth}
    \includegraphics[width=0.23\linewidth]{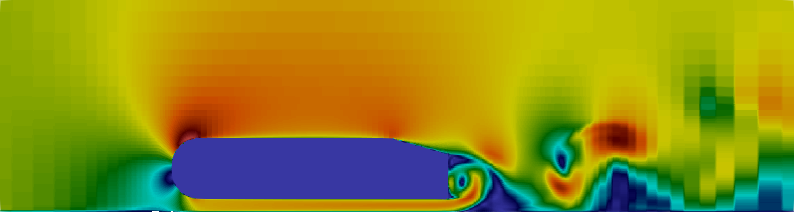}
    \caption{High fidelity training flow : slant angle $\alpha = 15.556^{\circ}$}
\end{subfigure}
\begin{subfigure}[b]{\textwidth}
    \centering
    \includegraphics[width=0.23\linewidth]{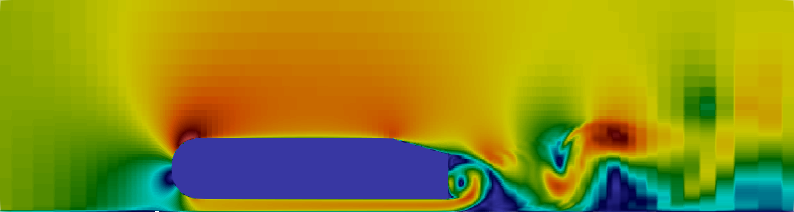}%
    \hspace*{0.01\linewidth}
    \includegraphics[width=0.23\linewidth]{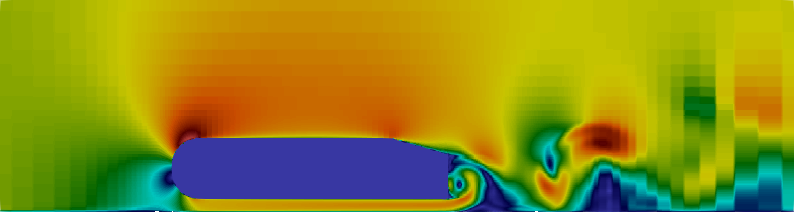}%
    \hspace*{0.01\linewidth}
    \includegraphics[width=0.23\linewidth]{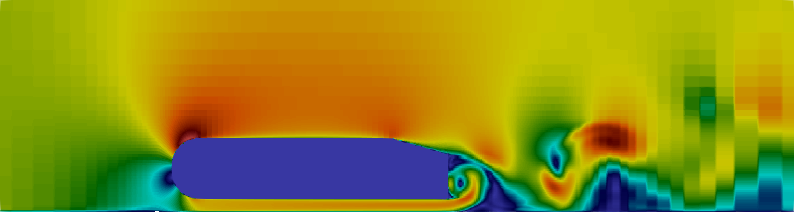}%
    \hspace*{0.01\linewidth}
    \includegraphics[width=0.23\linewidth]{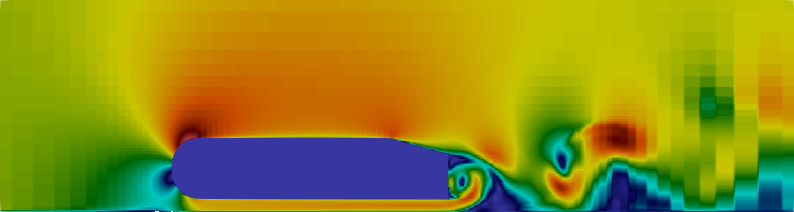}
    \caption{I-GILD}
\end{subfigure}

\caption{Comparison of the I-GILD and POD reconstructed flows alongside the high-fidelity flow for the training slant angle $\alpha = 15.556^{\circ}$, at time instants of $t=4.5$ s, $t=6.75$ s, $t=9.$ s, and $t=11.25$ s, respectively from left to right.}
\label{Fig. AB flow test 15.556}
\end{figure}

\begin{figure}[hbtp!]
\centering
\begin{subfigure}[b]{\textwidth}
    \centering
    \includegraphics[width=0.23\linewidth]{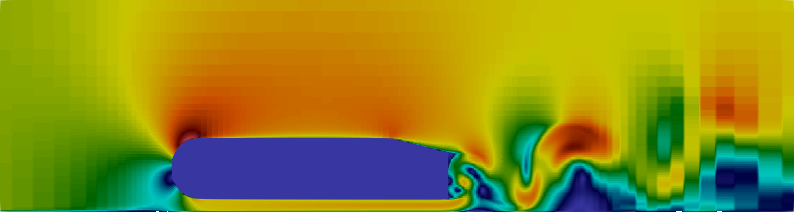}%
    \hspace*{0.01\linewidth}
    \includegraphics[width=0.23\linewidth]{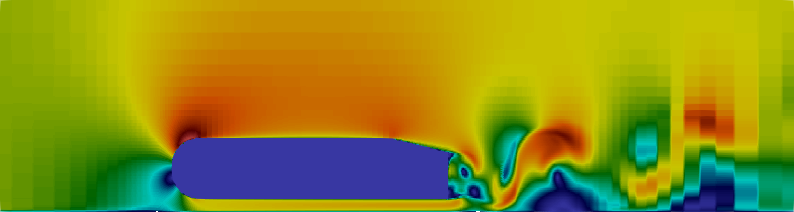}%
    \hspace*{0.01\linewidth}
    \includegraphics[width=0.23\linewidth]{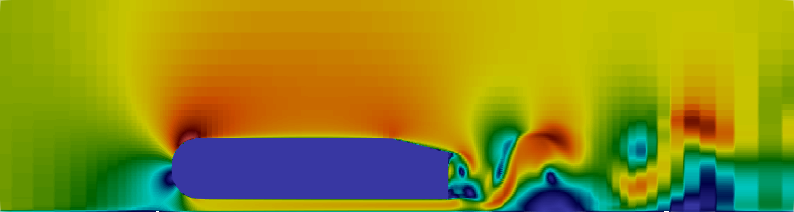}%
    \hspace*{0.01\linewidth}
    \includegraphics[width=0.23\linewidth]{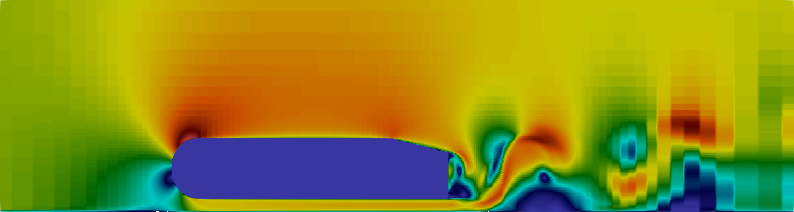}
    \caption{High fidelity unseen case : $\alpha = 9.944^{\circ}$}
\end{subfigure}
\begin{subfigure}[b]{\textwidth}
    \centering
    \includegraphics[width=0.23\linewidth]{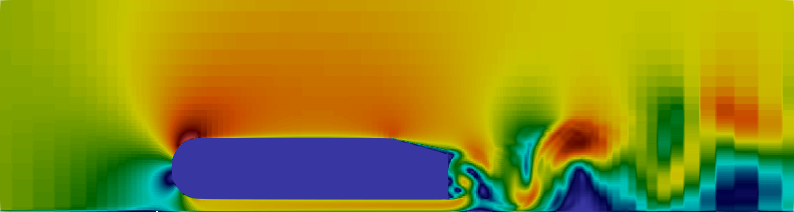}%
    \hspace*{0.01\linewidth}
    \includegraphics[width=0.23\linewidth]{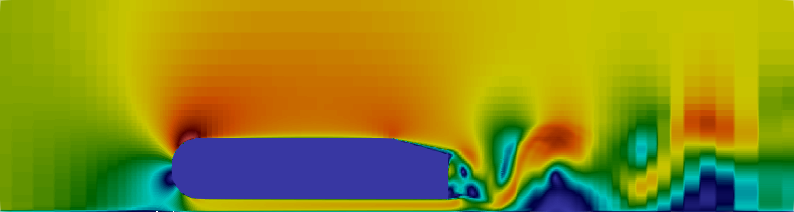}%
    \hspace*{0.01\linewidth}
    \includegraphics[width=0.23\linewidth]{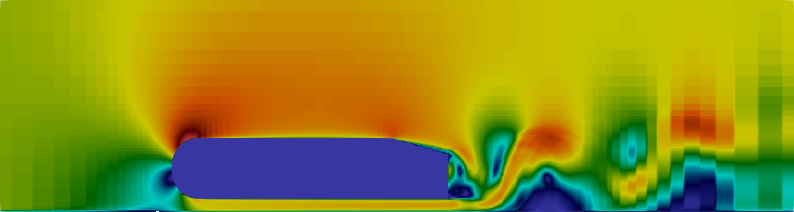}%
    \hspace*{0.01\linewidth}
    \includegraphics[width=0.23\linewidth]{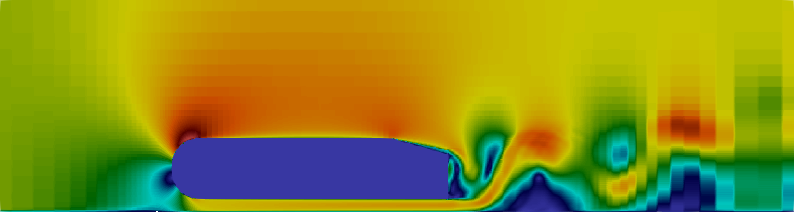}
    \caption{I-GILD}
\end{subfigure}

\caption{Comparison of the I-GILD and POD reconstructed flows alongside the high-fidelity flow for the test slant angle $\alpha = 9.944^{\circ}$, at time instants of $t=4.5$ s, $t=6.75$ s, $t=9.$ s, and $t=11.25$ s, respectively from left to right.}
\label{Fig. AB flow test 9.944}
\end{figure}

\begin{figure}[hbtp!]
\centering
\begin{subfigure}[b]{\textwidth}
    \centering
    \includegraphics[width=0.23\linewidth]{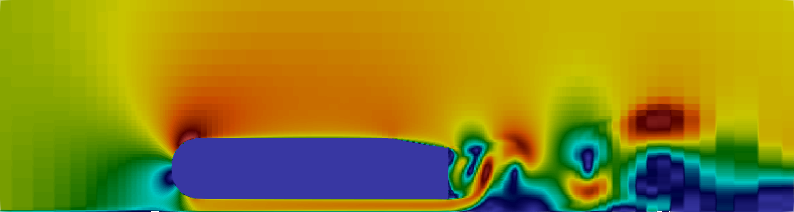}%
    \hspace*{0.01\linewidth}
    \includegraphics[width=0.23\linewidth]{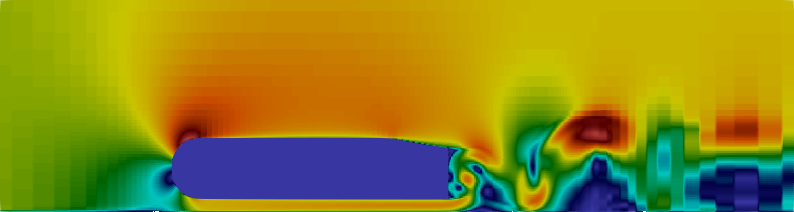}%
    \hspace*{0.01\linewidth}
    \includegraphics[width=0.23\linewidth]{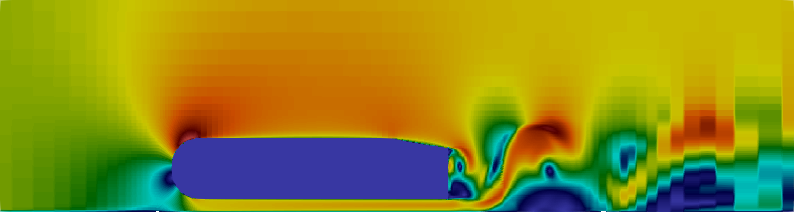}%
    \hspace*{0.01\linewidth}
    \includegraphics[width=0.23\linewidth]{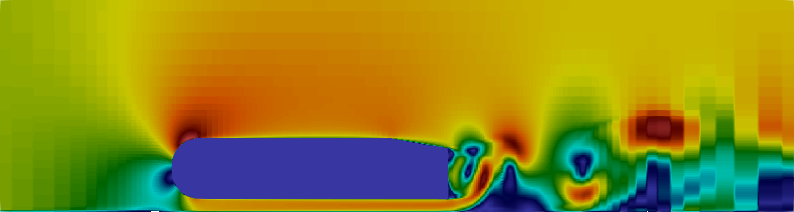}
    \caption{High fidelity unseen case : $\alpha = 14.085^{\circ}$}
\end{subfigure}
\begin{subfigure}[b]{\textwidth}
    \centering
    \includegraphics[width=0.23\linewidth]{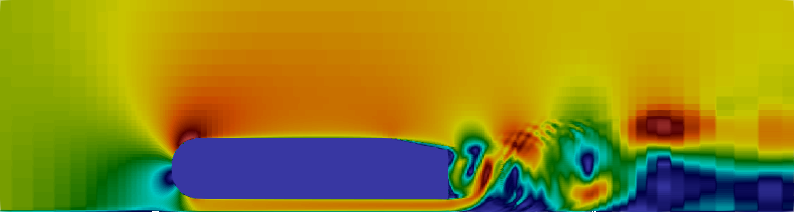}%
    \hspace*{0.01\linewidth}
    \includegraphics[width=0.23\linewidth]{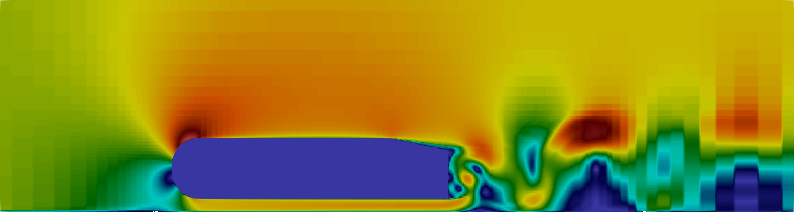}%
    \hspace*{0.01\linewidth}
    \includegraphics[width=0.23\linewidth]{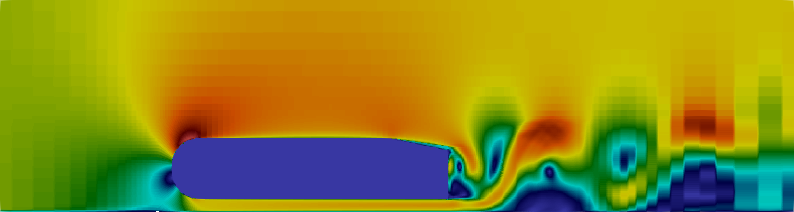}%
    \hspace*{0.01\linewidth}
    \includegraphics[width=0.23\linewidth]{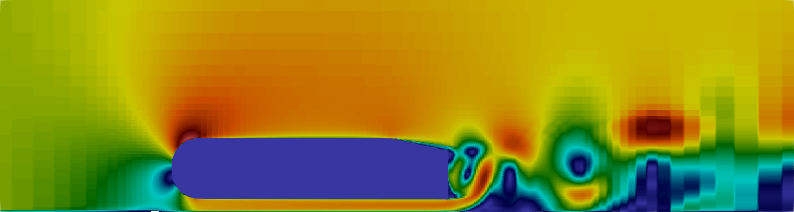}
    \caption{I-GILD}
\end{subfigure}

\caption{Comparison of the IGILD and POD reconstructed flows alongside the high-fidelity flow for the test slant angle $\alpha = 14.085^{\circ}$, at time instants of $t=4.5$ s, $t=6.75$ s, $t=9$ s, and $t=11.25$ s, respectively from left to right.}
\label{Fig. AB flow test 14.085}
\end{figure}

Regarding error growth over time, I-GILD consistently outperforms GILD, as illustrated in Figure \ref{Fig. Errors wrt time Ahmed Body}. For the training slant angles, I-GILD closely matches the POD error within the seen time intervals, confirming its ability to accurately capture the dominant flow features. Moreover, in the unseen time intervals, I-GILD demonstrates strong generalization capabilities, maintaining high accuracy even when extrapolating to regions where it was not explicitly trained. In contrast, GILD's poor performance is more noticeable across time.
For the test slant angles $ 9.944^{\circ} $ and $ 14.085^{\circ} $, I-GILD continues to exhibit better accuracy than GILD. Although both models perform reasonably well, I-GILD's superior accuracy across all time steps, including unseen slant angles, further reinforces its robustness.
This is further confirmed when inspecting the reconstructed flow fields, shown in Figures \ref{Fig. AB flow test 7.998}, \ref{Fig. AB flow test 11.848}, and \ref{Fig. AB flow test 15.556} for the training angles, and Figures \ref{Fig. AB flow test 9.944} and \ref{Fig. AB flow test 14.085} for the test angles. Notably, I-GILD reconstruction accurately captures the wake structure behind the Ahmed body by closely mimicking the flow patterns observed in the high-fidelity model, successfully capturing the flow separation at the start of the slant surface and its reattachment near the end.
 
In summary, I-GILD demonstrates superior learning performance over the original GILD method, achieving faster convergence, better residual reduction, and improved accuracy for both training and test angles.

{
\subsection{Lid-Driven Cylindrical Cavity with Variable Reynolds Number}

This study evaluates the performance of the I-GILD algorithm in predicting two-dimensional flow within a lid-driven cylindrical cavity using entirely experimental data. Working with experimental measurements introduces challenges such as measurement noise and slight inconsistencies, thus testing the model's robustness and ability to extend to real-world conditions beyond synthetic or simulation-based datasets.

The flow is characterized by varying Reynolds numbers, achieved by adjusting the characteristic velocity $U$ while keeping the geometry and fluid properties constant. Higher Reynolds numbers correspond to increased velocities and more complex flow regimes. In this example, four Reynolds numbers were chosen for training: $Re = 4614$, $Re = 4704$, $Re = 4885$, and $Re = 4976$, representing different stages of the flow and exposing the I-GILD model to a wide range of flow dynamics, including structure transitions and vortex formations. A fifth Reynolds number, $Re = 4795$, not included in the training set, was used for testing to assess the model's ability to generalize to unseen flow conditions.

The dataset consisting of velocity measurements were obtained from \cite{Schmid2009}, where a typical PIV setup was used as depicted in Figure~\ref{Fig. geometry_cavity}. It comprises $200$ experimental velocity snapshots per Reynolds number, capturing the temporal evolution of the flow field. POD was performed on all $200$ snapshots to ensure that the dominant flow structures were adequately captured. A total of $60$ modes were retained to maintain computational efficiency while capturing essential flow features.

For training the I-GILD model, the first $70\%$ of the snapshots ($140$ time steps) were used, while the remaining $30\%$ ($60$ snapshots) were reserved for testing in unseen forecast time. This allowed us to assess the model's predictive performance both within the training interval and in extrapolation to future time steps. Additionally, we tested on the Reynolds number $Re = 4795$, which was not included in the training set, in order to examine the model's ability to generalize across different flow conditions.

\begin{figure}[hbtp!]
\centering
\includegraphics[width=0.4\textwidth]{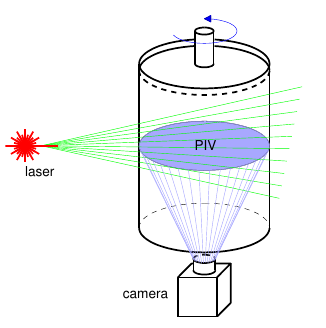}
\caption{Sketch of geometry and experimental setup of flow in a lid-driven cylindrical cavity.}
\label{Fig. geometry_cavity}
\end{figure}

In our evaluation of the I-GILD model's predictive capabilities, we first analyzed the error metrics associated with the model's predictions in comparison to high-fidelity data obtained from PIV experiments. For all the training Reynolds numbers, Figure \ref{Fig. cavity error train 1}, Figure \ref{Fig. cavity error train 2}, Figure \ref{Fig. cavity error train 3} and Figure\ref{Fig. cavity error train 4} show that the I-GILD model exhibits strong agreement with the POD during the training time interval. This is evidenced by minimal error levels, closely resembling those observed in the POD, indicating that the model effectively captures the underlying flow dynamics. In the subsequent temporal extrapolation interval, the I-GILD model continues to perform remarkably, achieving error levels that remain low, typically around a $2\%$ deviation from the POD at the worst case scenario.

Figure~\ref{Fig. cavity error untrain} further demonstrates the model's robust generalization capabilities, particularly for the Reynolds number $Re = 4795$, which was not part of the training dataset, with a mean error deviation of around $7\%$ with respect to POD. This result highlights the I-GILD model’s capability to generalize across flow conditions not explicitly present during training.

To gain further and deeper insights into the model's predictive capabilities, we examine the reconstructed flow fields, as shown in Figure~\ref{Fig. washing machine training} and Figure~\ref{Fig. washing machine test}. For each training Reynolds number, we present visualizations of the flow fields at the final time step, which lies within the extrapolation temporal zone, outside the model’s training conditions. These results demonstrate that the I-GILD model successfully reconstructs flow structures that are well-aligned with the high-fidelity data, capturing essential flow dynamics even in regions beyond the training data. 
For the test case at $Re=4795$, a more challenging scenario due to its exclusion from the training dataset, the model is tasked with predicting in both time and parameter space. Despite this increased complexity, the I-GILD model yields results that remain in good agreement with the high-fidelity data, capturing the dominant flow structures across both untrained temporal and parameter domains. Some smoothness is observed in the predicted flow fields, which can be attributed to the model's difficulty in capturing the turbulent dynamics of low-amplitude modes. These modes typically contribute finer details to the flow structures, and their representation often requires a high level of sensitivity. While it remains a challenge for reduced-order models to accurately reconstruct these low-amplitude features, the I-GILD model's ability to approximate the dominant dynamics effectively suggests a promising predictive power and a pleasing robustness in handling high-dimensional, parameterized flows.

In summary, both the training and test results highlight the I-GILD model’s strength in generalizing to unseen conditions. The close match between the reconstructed flow fields and the high-fidelity data across different Reynolds numbers supports the model's efficacy in approximating complex flow dynamics, with minimal deviation in essential structures, even when extended beyond the conditions of training.

\begin{figure}[hbtp!]
\centering
\begin{subfigure}{0.32\textwidth}
    \includegraphics[width=\textwidth]{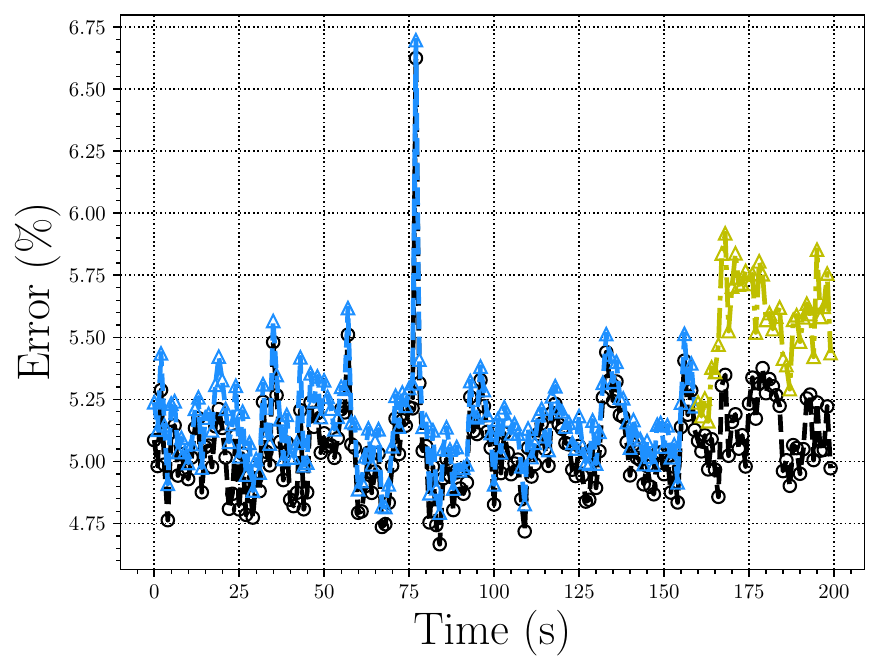}
    \caption{training $Re = 4614$}\label{Fig. cavity error train 1}
\end{subfigure}%
\begin{subfigure}{0.32\textwidth}
    \includegraphics[width=\textwidth]{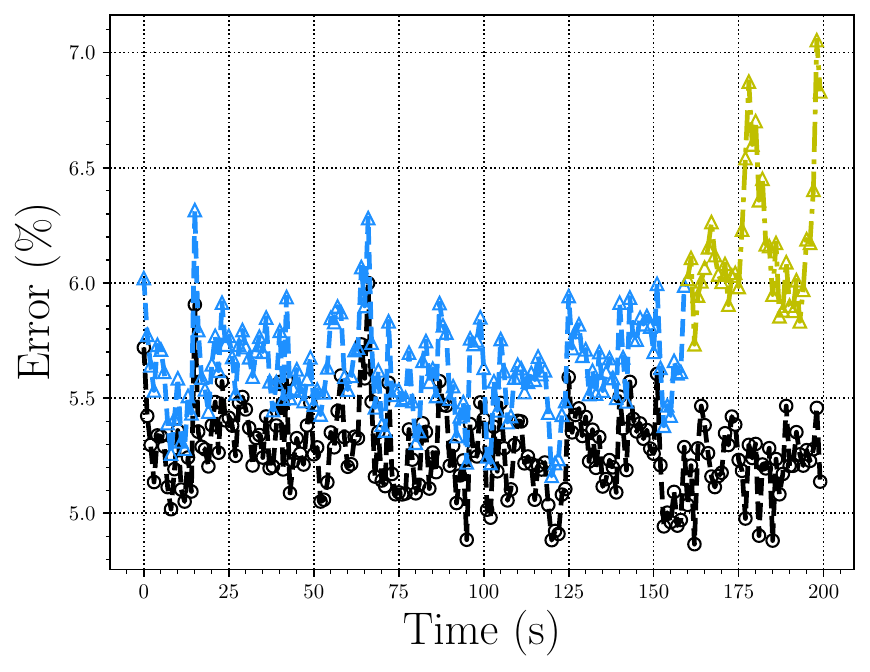}
    \caption{training $Re=4704$}\label{Fig. cavity error train 2}
\end{subfigure}%
\begin{subfigure}{0.32\textwidth}
    \includegraphics[width=\textwidth]{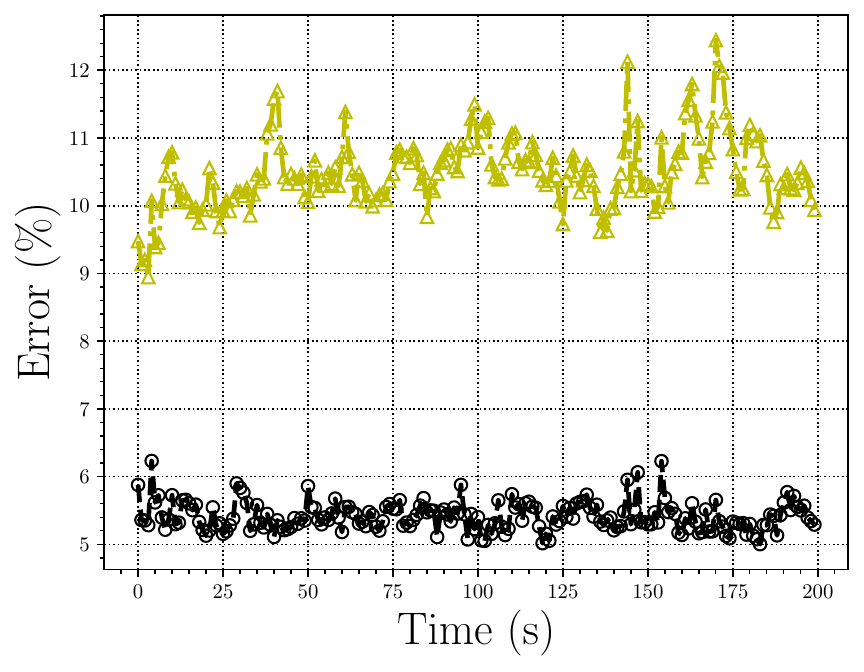}
    \caption{test $Re=4795$}\label{Fig. cavity error train 3}
\end{subfigure}

\begin{subfigure}{0.32\textwidth}
    \includegraphics[width=\textwidth]{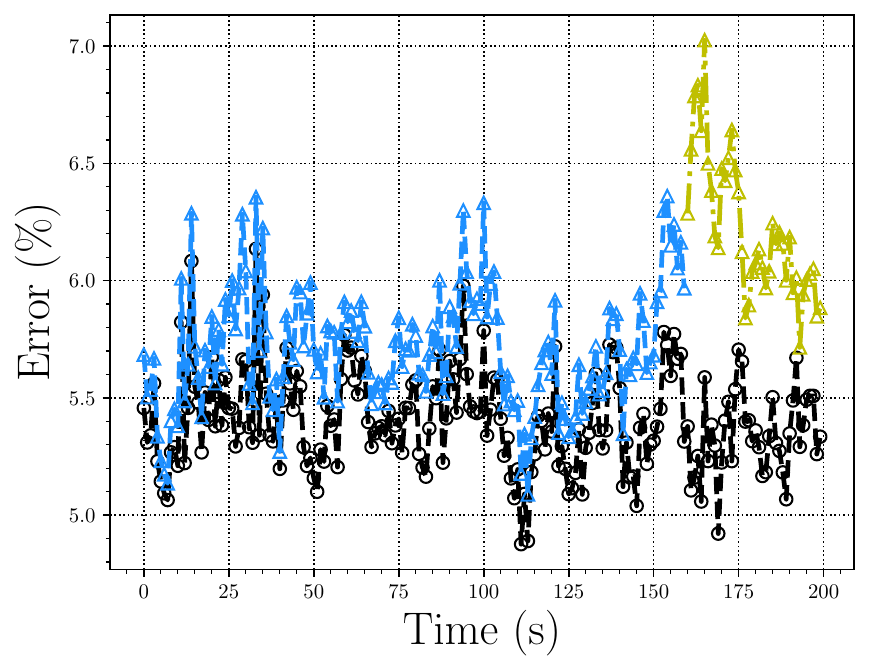}
    \caption{training $Re=4885$}\label{Fig. cavity error train 4}
\end{subfigure}%
\begin{subfigure}{0.32\textwidth}
    \includegraphics[width=\textwidth]{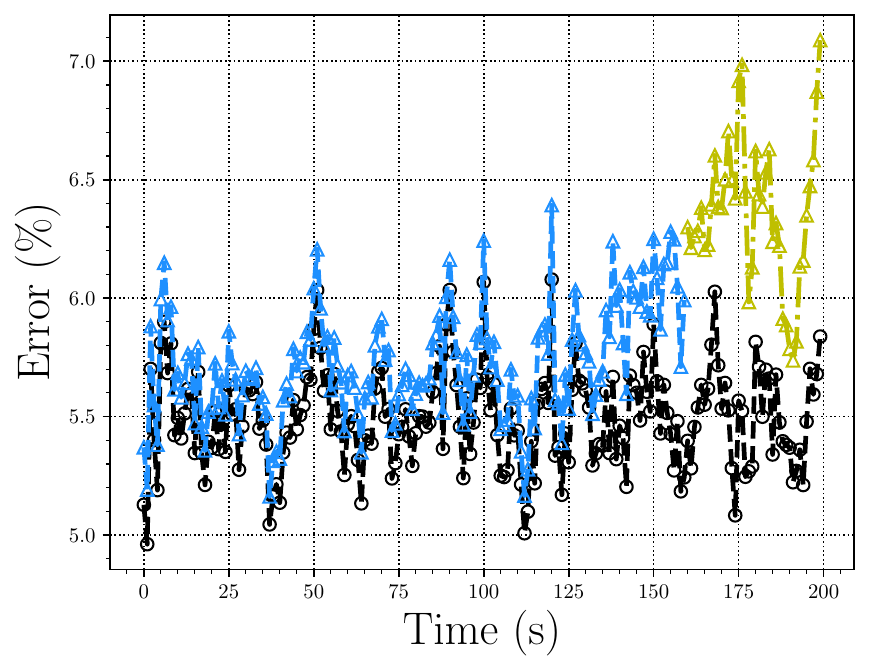}
    \caption{training $Re=4976$}\label{Fig. cavity error untrain}
\end{subfigure}

\includegraphics[width=0.5\textwidth]{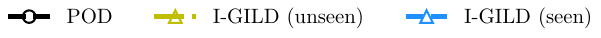}
\caption{Percentage of error with respect to time for the training and test Reynolds numbers.}
\label{Fig. Errors wrt time washin machine}
\end{figure}

\begin{figure}[hbtp!]
\centering
\begin{subfigure}[b]{\textwidth}
    \centering
    \hspace*{0.01\linewidth}
    \includegraphics[width=0.2\linewidth]{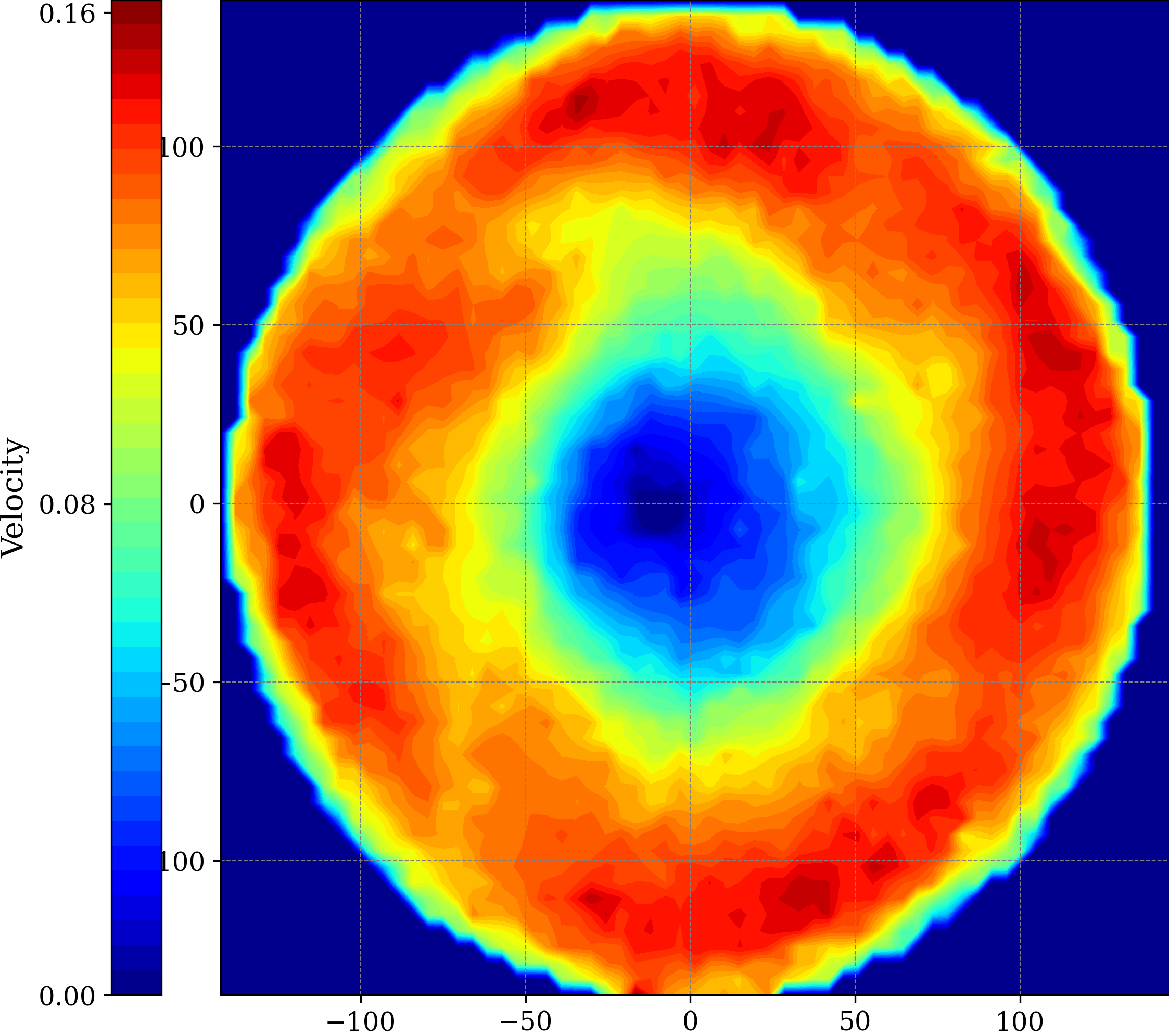}%
    \hspace*{0.01\linewidth}
    \includegraphics[width=0.2\linewidth]{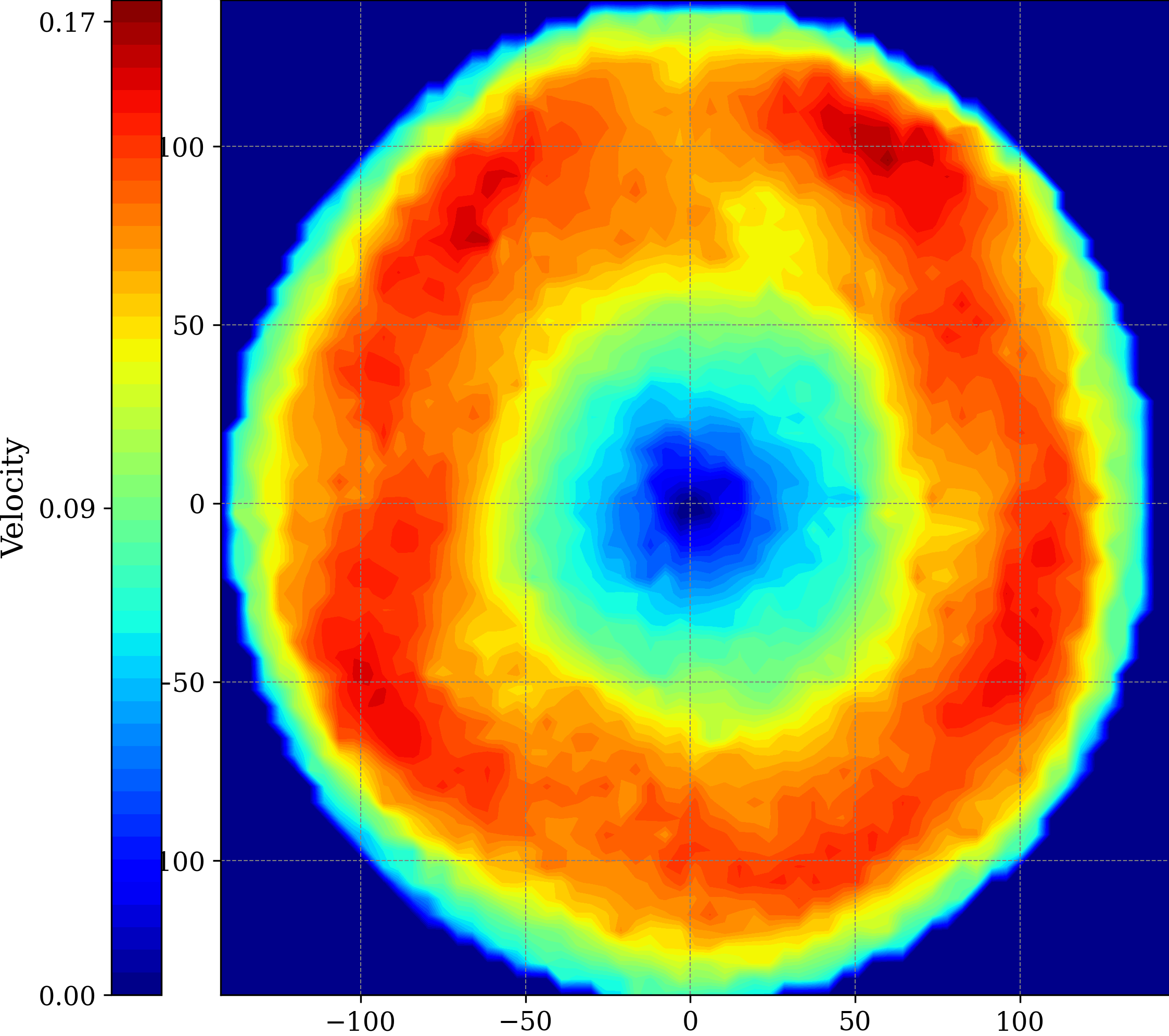}
    \hspace*{0.01\linewidth}
    \includegraphics[width=0.2\linewidth]{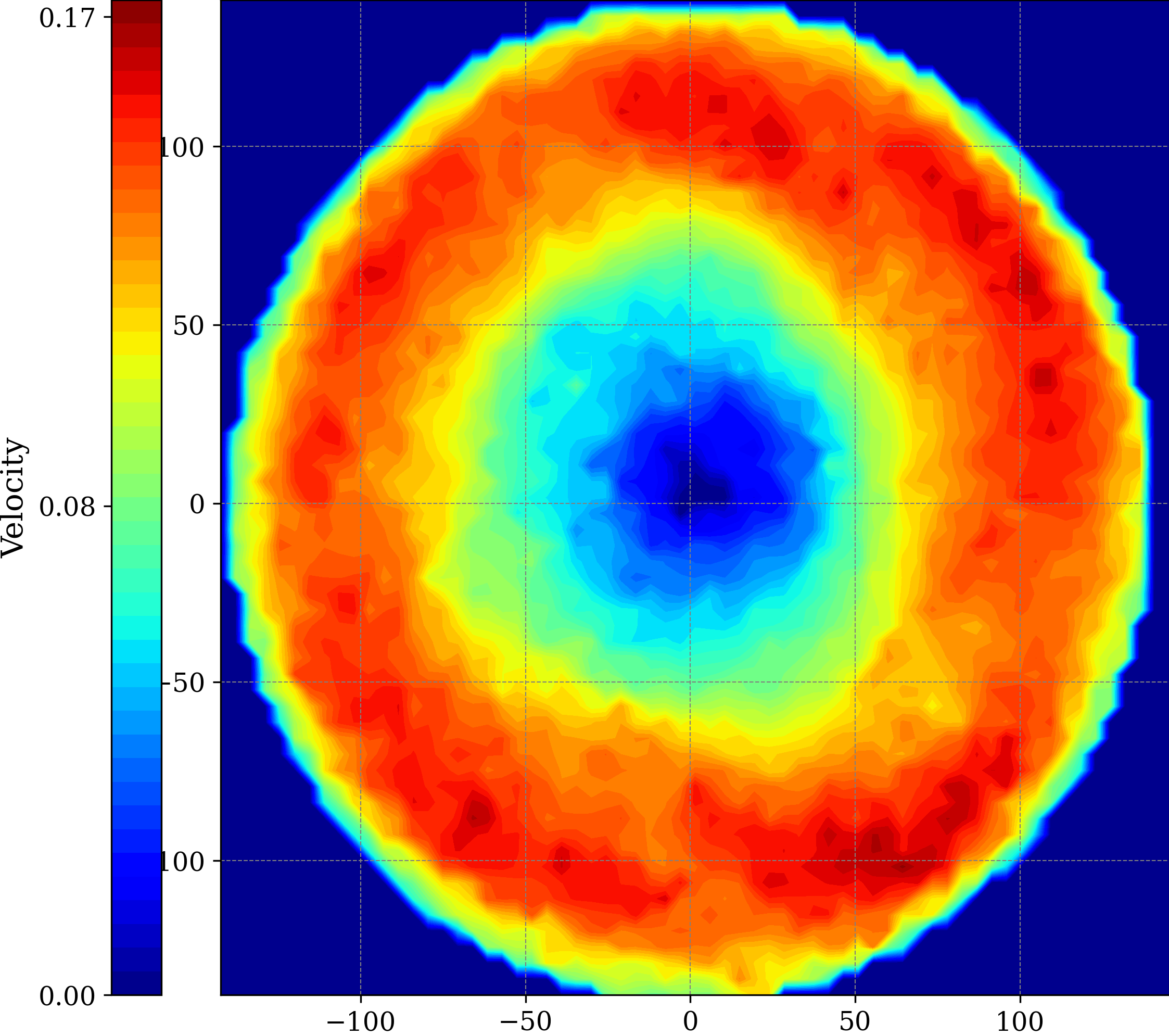}
    \hspace*{0.01\linewidth}
    \includegraphics[width=0.2\linewidth]{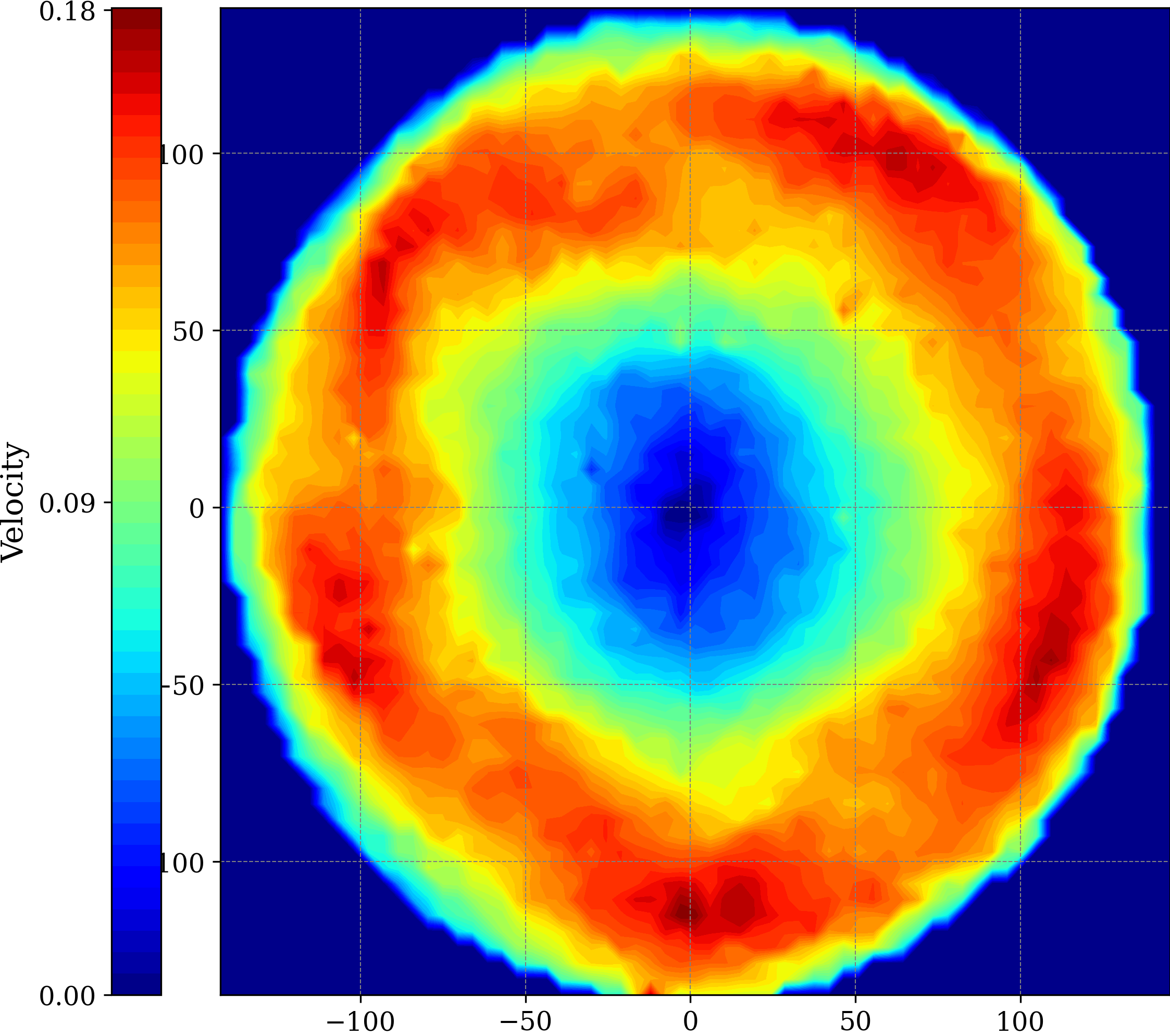}
    \caption{Ground truth flows}
\end{subfigure}
\\
\begin{subfigure}[b]{\textwidth}
    \centering
    \hspace*{0.01\linewidth}
    \includegraphics[width=0.2\linewidth]{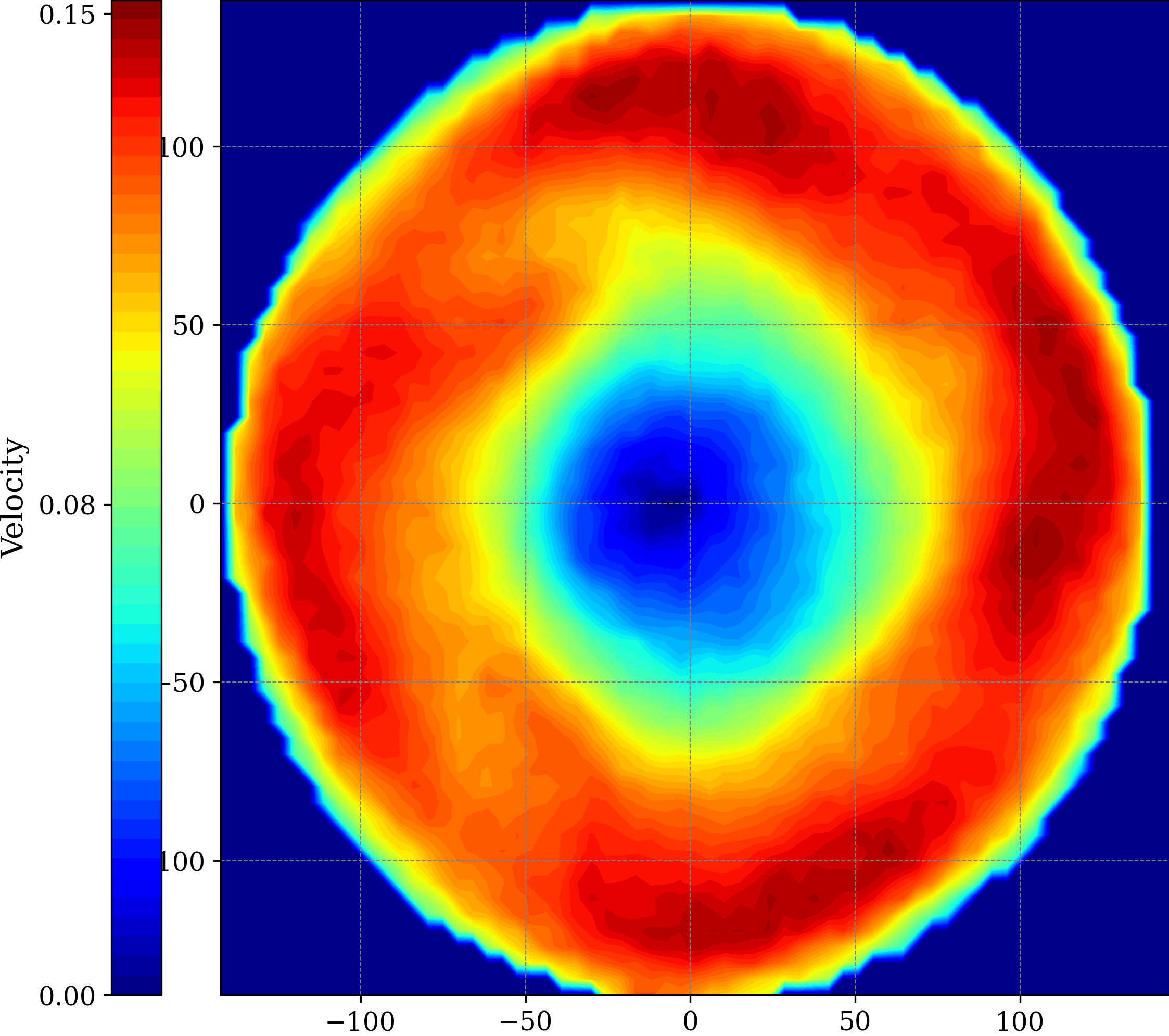}%
    \hspace*{0.01\linewidth}
    \includegraphics[width=0.2\linewidth]{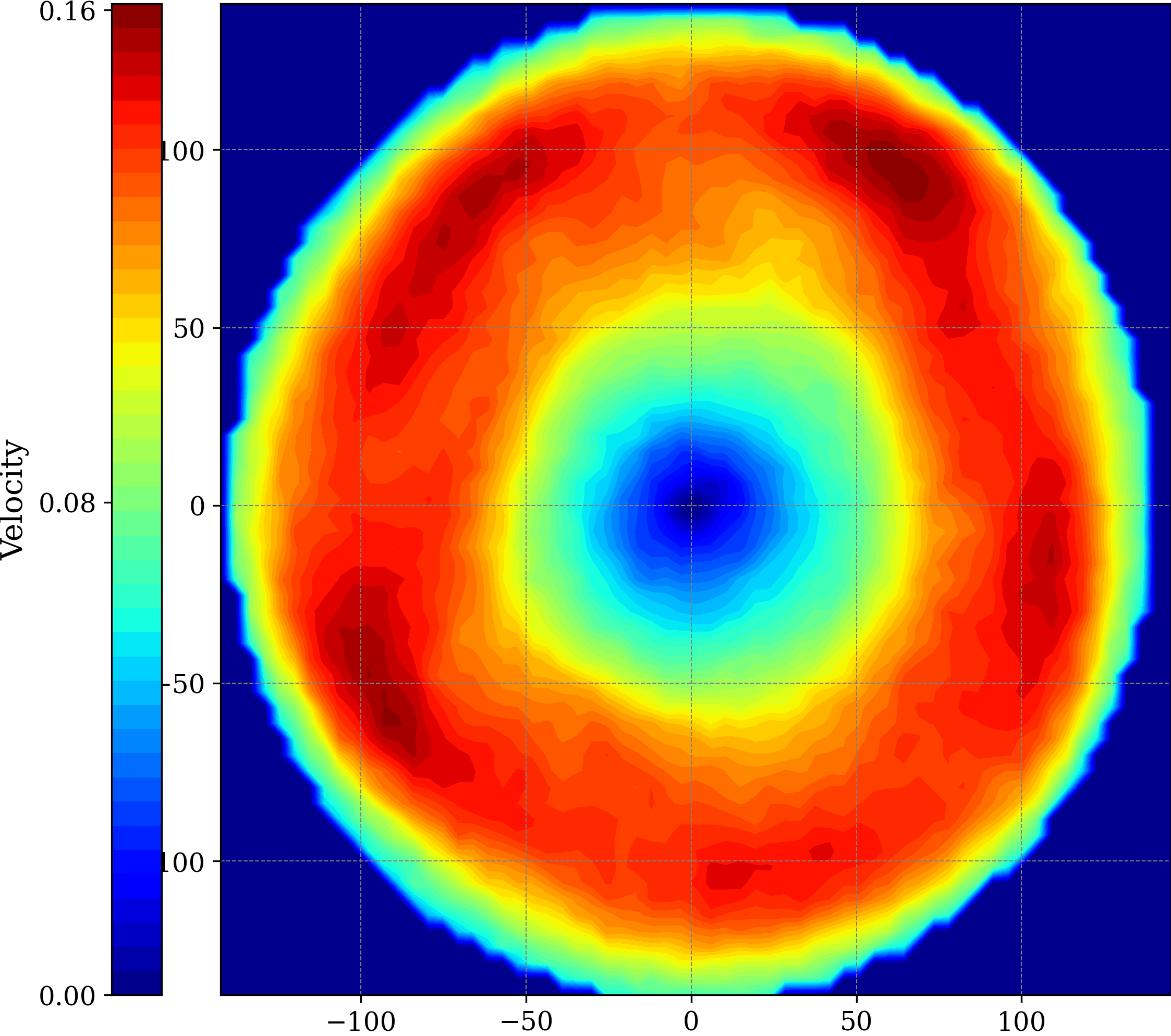}
    \hspace*{0.01\linewidth}
    \includegraphics[width=0.2\linewidth]{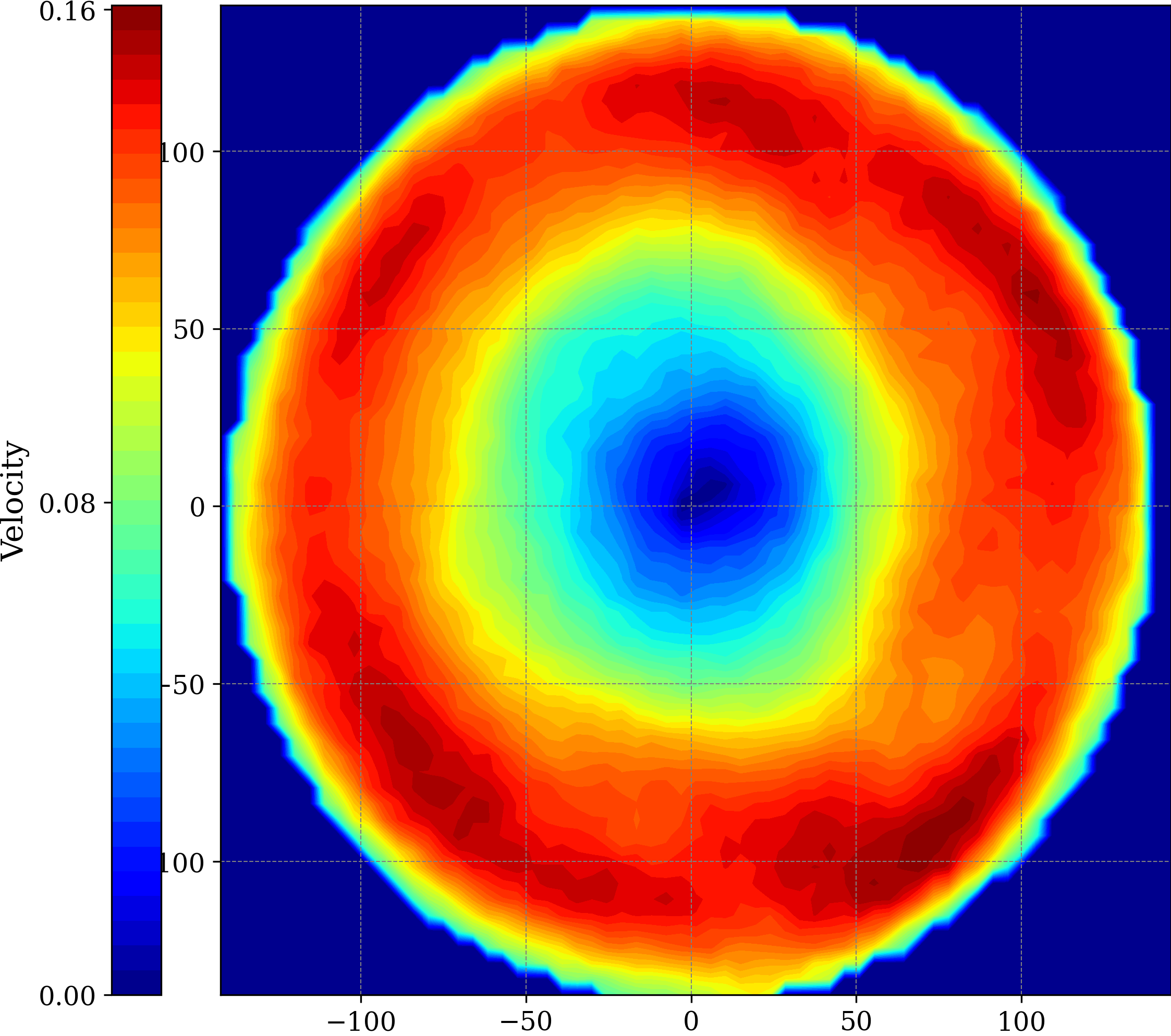}
    \hspace*{0.01\linewidth}
    \includegraphics[width=0.2\linewidth]{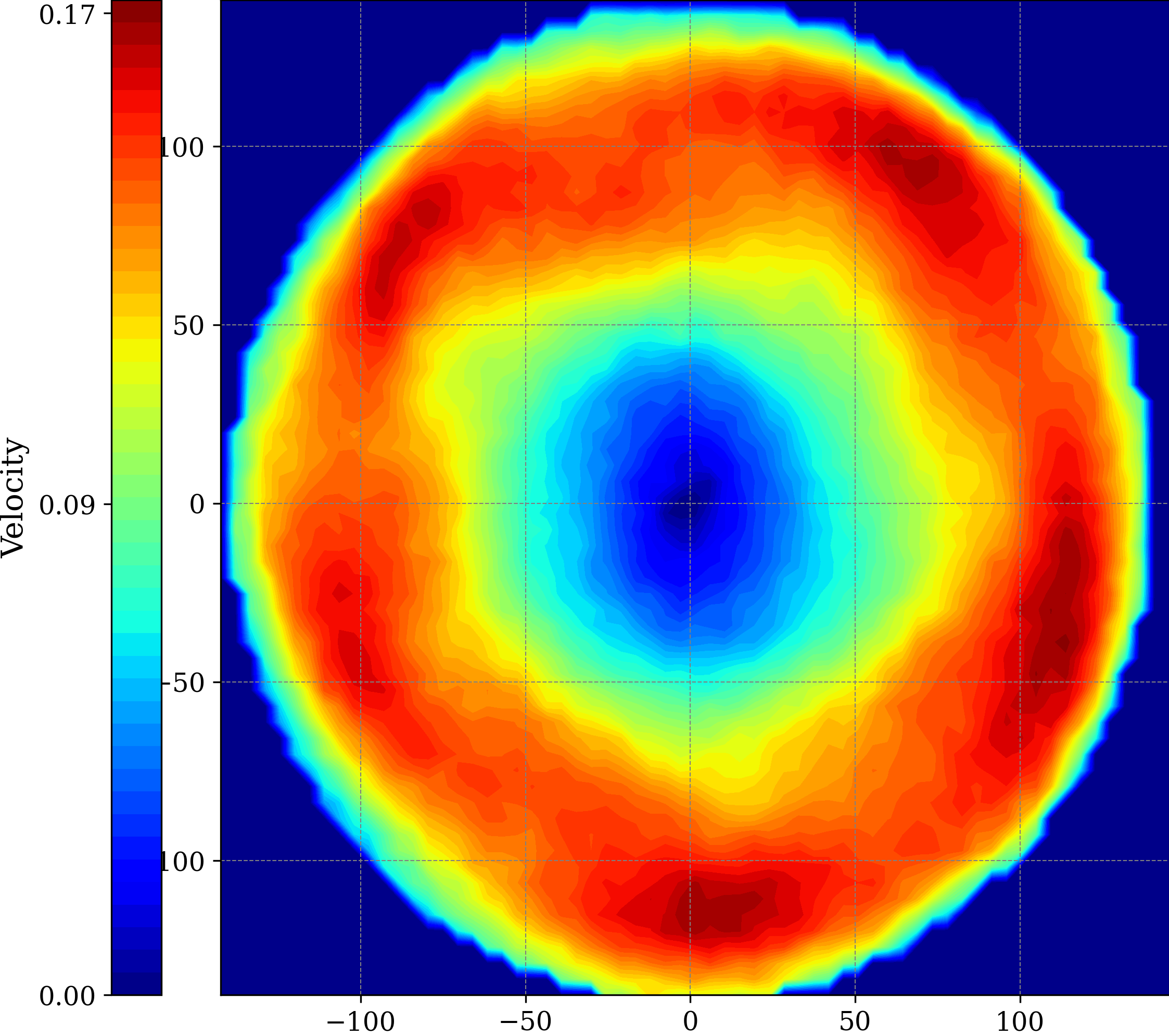}
    \caption{POD flows}
\end{subfigure}
\\
\begin{subfigure}[b]{\textwidth}
    \centering
    \hspace*{0.01\linewidth}
    \includegraphics[width=0.2\linewidth]{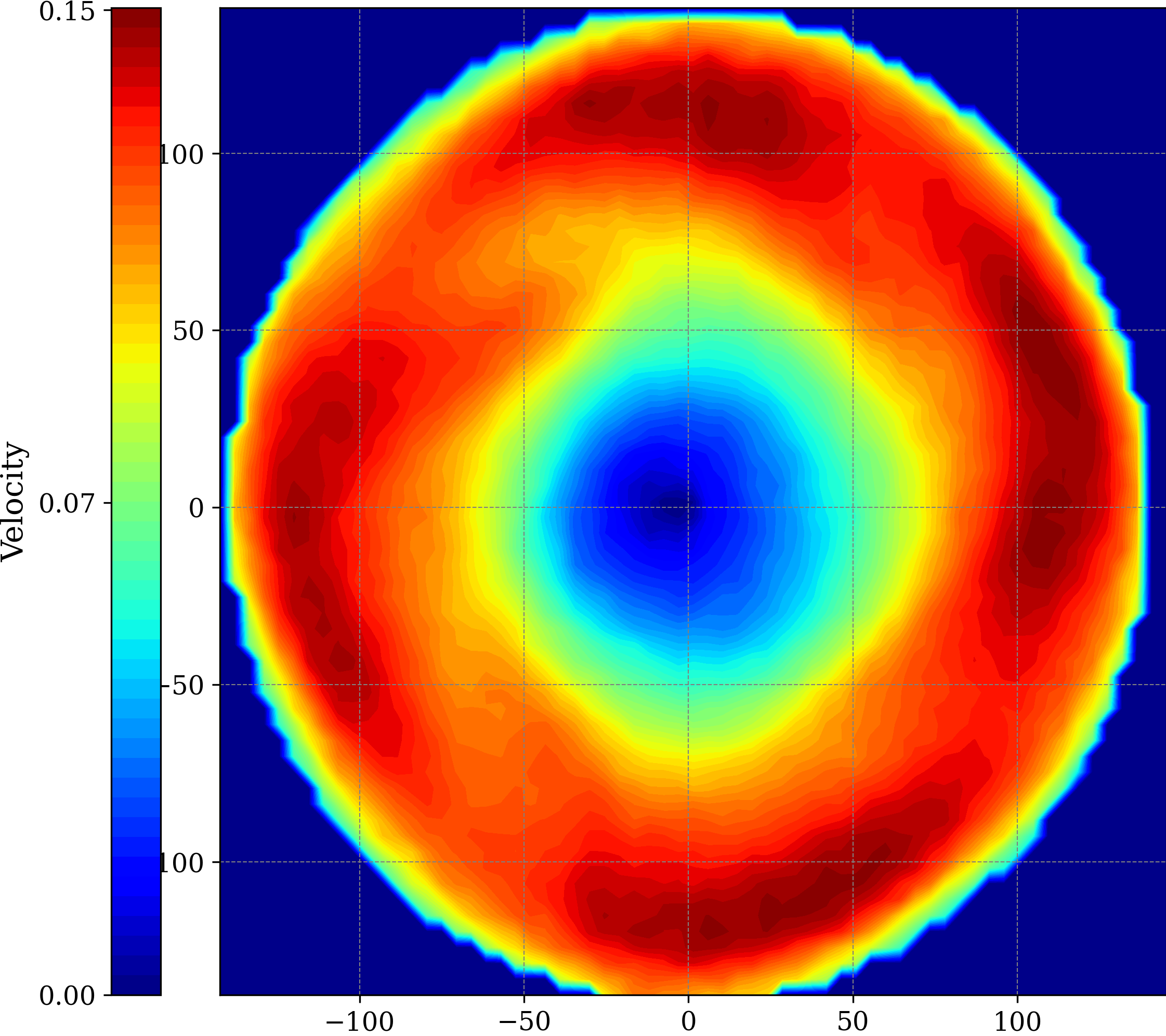}%
    \hspace*{0.01\linewidth}
    \includegraphics[width=0.2\linewidth]{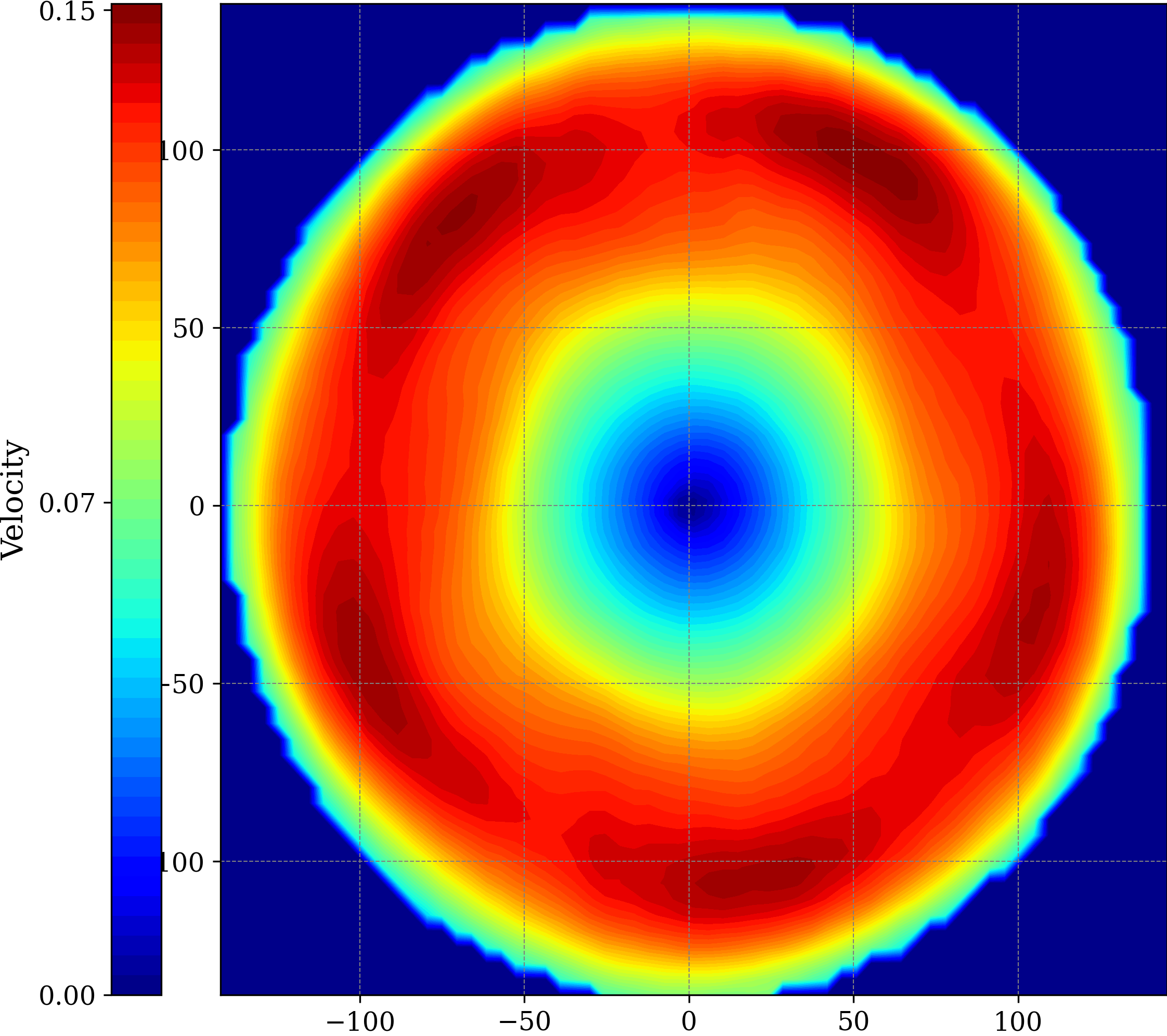}
    \hspace*{0.01\linewidth}
    \includegraphics[width=0.2\linewidth]{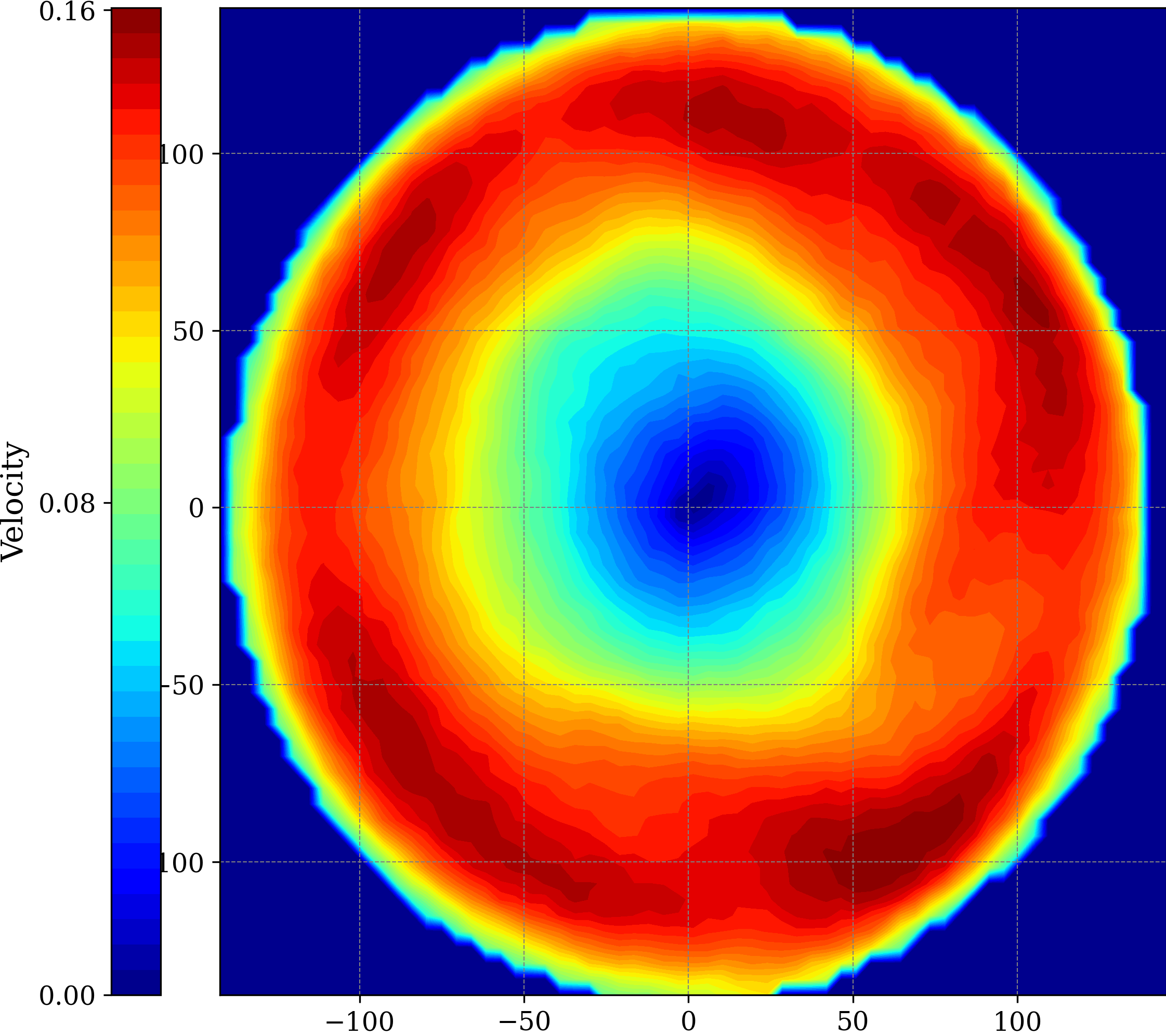}
    \hspace*{0.01\linewidth}
    \includegraphics[width=0.2\linewidth]{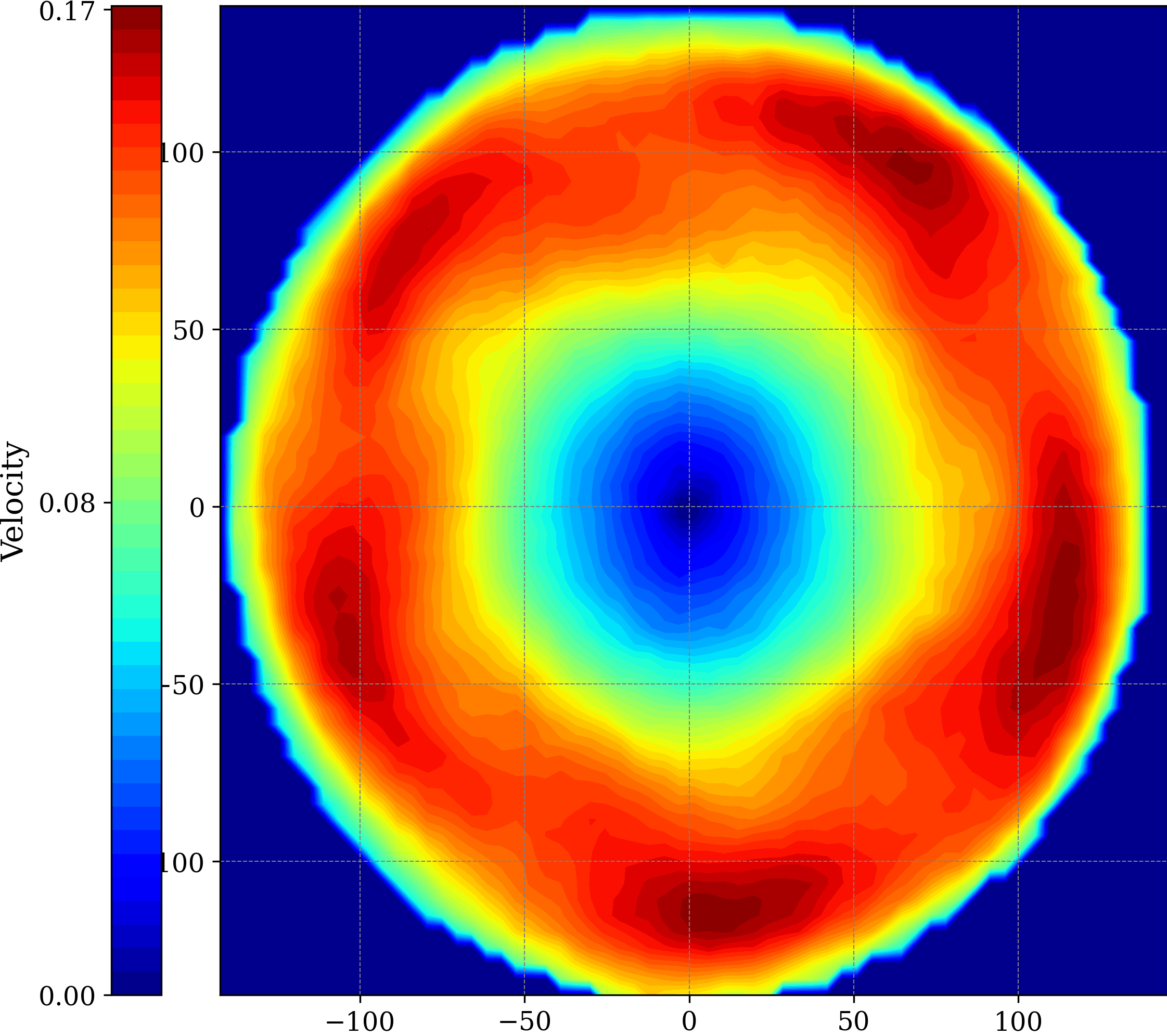}
    \caption{I-GILD predictions}
\end{subfigure}
\caption{Comparison of I-GILD reconstructed flow fields with POD and high-fidelity flow fields at the final time instant for the training Reynolds numbers: 4614, 4795, 4885, and 4976, respectively, from left to right.}
\label{Fig. washing machine training}
\end{figure}

\begin{figure}[hbtp!]
\centering
\begin{subfigure}[b]{0.2\textwidth}
    \centering
    \includegraphics[width=\linewidth]{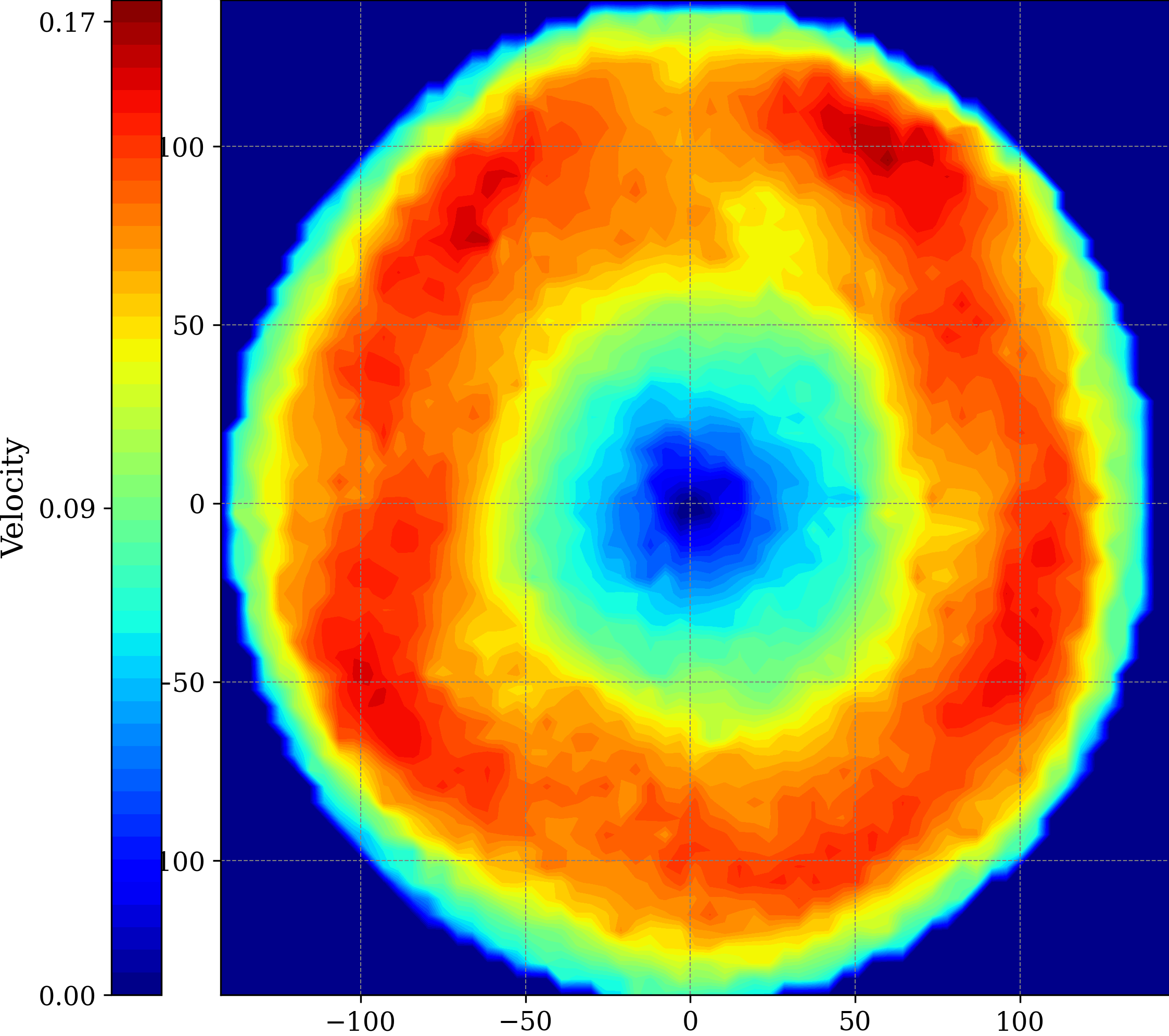}
    \caption{Gound truth}
\end{subfigure}%
\begin{subfigure}[b]{0.2\textwidth}
    \centering
    \includegraphics[width=\linewidth]{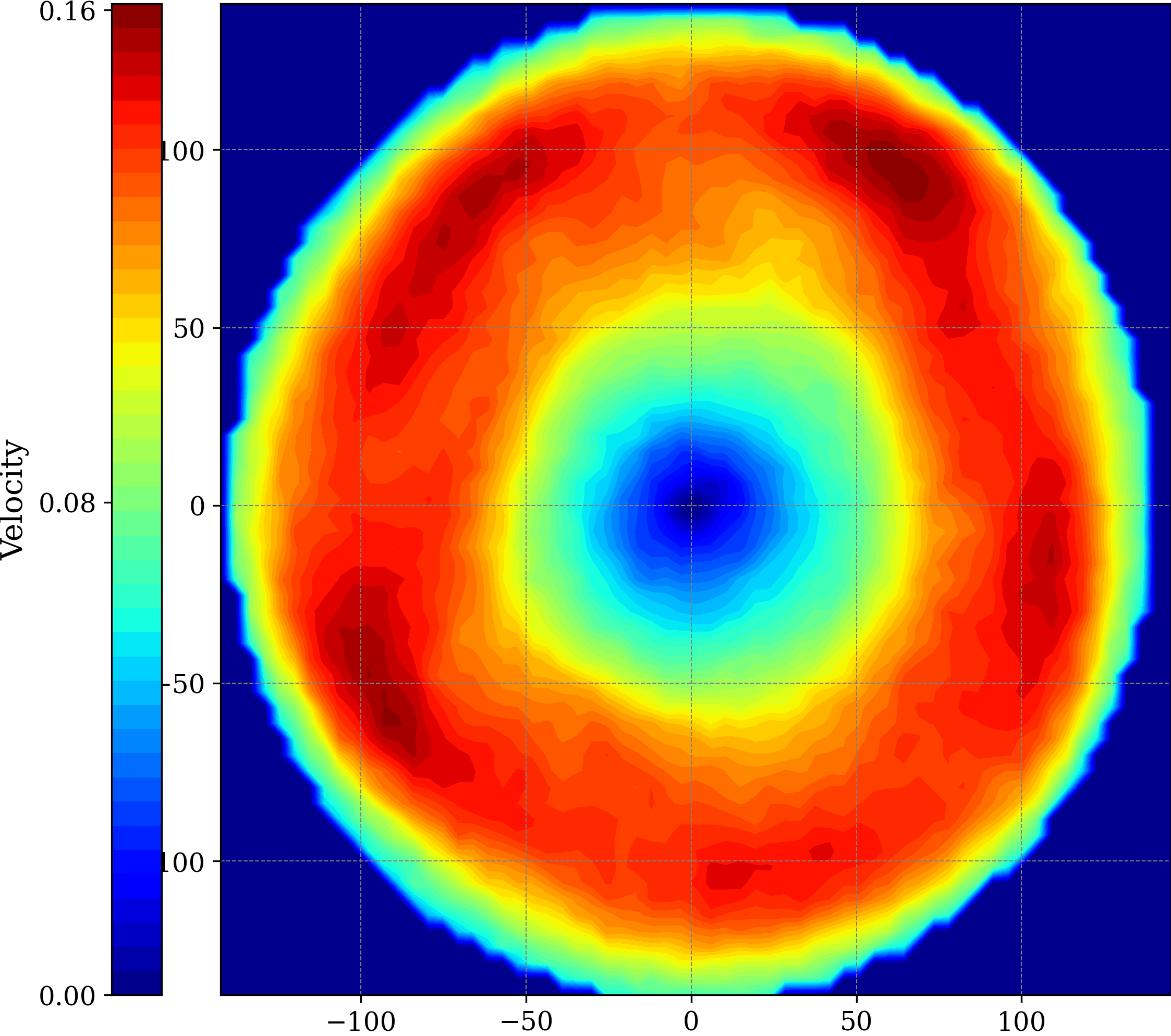}
    \caption{POD}
\end{subfigure}%
\begin{subfigure}[b]{0.2\textwidth}
    \centering
    \includegraphics[width=\linewidth]{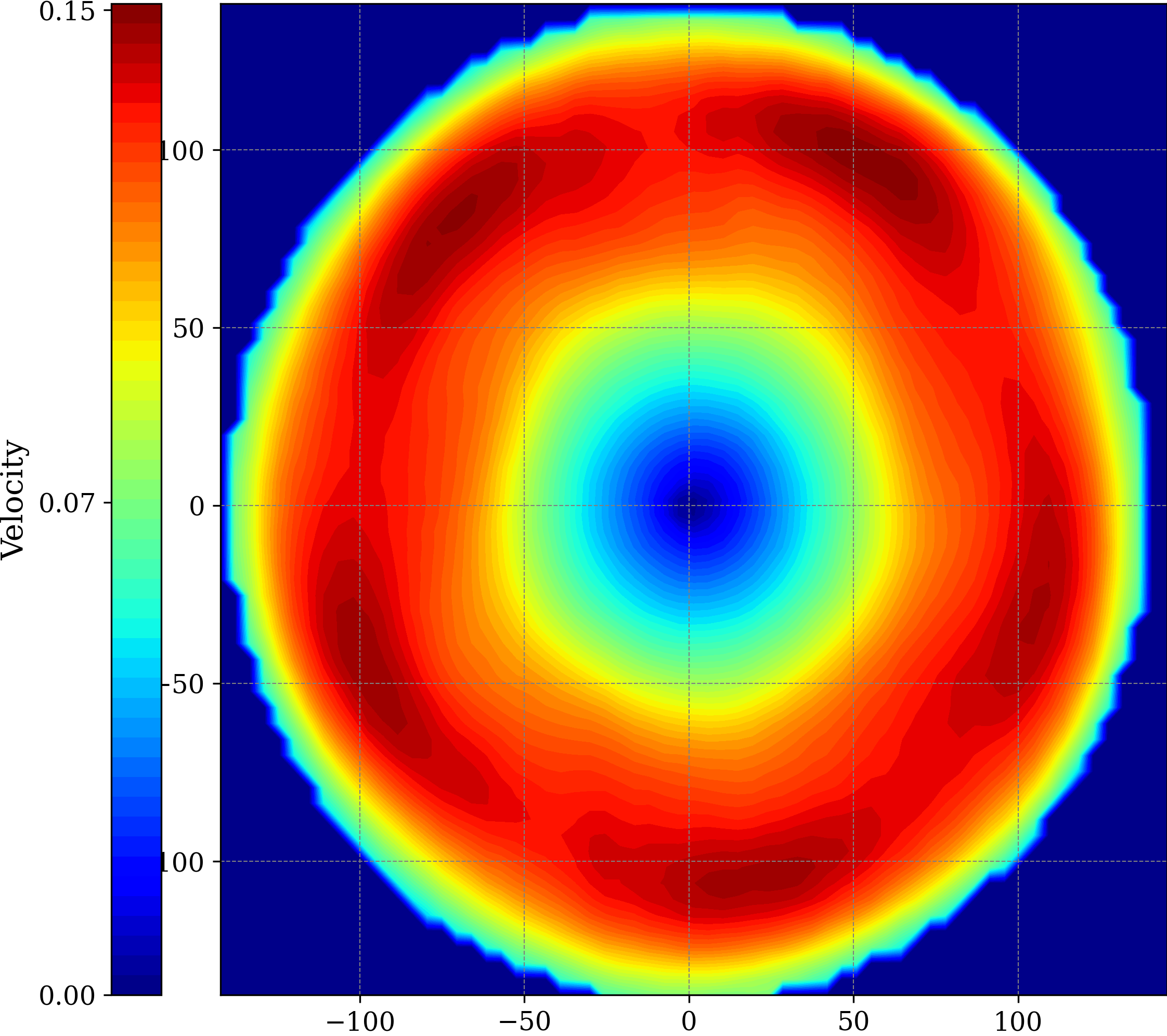}
    \caption{I-GILD}
\end{subfigure}

\caption{Comparison of I-GILD-reconstructed flow fields with POD and high-fidelity flow fields at the final time instant for the test Reynolds number 4795.}
\label{Fig. washing machine test}
\end{figure}

\section{Summary and conclusions}

In this study, we present the Improved Greedy Identification of Latent Dynamics (I-GILD) method, demonstrating robustness and computational efficiency in managing high-dimensional systems derived from diverse data sources. Our findings reveal that I-GILD accelerates convergence, significantly reducing computational effort and achieving low learning residuals.

Building on the foundation of GILD method \cite{GILD2024}, I-GILD reformulates the minimization problem using the Frobenius norm and employs the conjugate-gradient (CG) method to efficiently solve the underlying generalized Sylvester problem. Additionally, the derived error bounds provide strong assurances of reliability and predictive accuracy, assuming bounded spectral norms of the identified operators, low learning residuals and low initial error, thereby reinforcing the method's theoretical rigor.

Our numerical experiments in cases such as Ahmed body flow and the cylindrical lid-driven cavity affirm that I-GILD achieves a superior representation of the latent dynamics. These results emphasize the importance of advancing model reduction techniques that balance computational efficiency with fidelity in capturing key predictive features. In experiments with real-world data from the lid-driven cylindrical cavity, the method demonstrates robustness under conditions of measurement noise and slight data inconsistencies. It generalizes effectively across varying Reynolds numbers and reconstructs flow fields that are closely aligned with high-fidelity data, confirming the method's capability to manage high-dimensional, parameterized flows under real-world conditions.

In summary, the advancements introduced in I-GILD represent a promising step forward in data-driven reduced-order modeling for complex systems, bridging the gap between computational efficiency and high accuracy. The method’s non-intrusive, data-driven character, its convergence properties, and rigorous validation, position I-GILD as an essential tool for researchers and engineers working with dynamical systems with  second order nonlinearities. 

{
Future work will focus on incorporating physics constraints within the latent dynamics to further enhance the stability and interpretability of reduced-order models. Specifically, ensuring that the linear term of the latent-space dynamics symmetric positive (or negative) definite or simply symmetric and the quadratic term is skew-symmetric introduces meaningful physical properties to the system. The quadratic term, being skew-symmetric, ensures that it does not introduce or dissipate energy in the system, thus contributing to stability without influencing the overall energy balance. On the other hand, the linear term, as a symmetric operator, governs dissipation or growth in the system, acting as a stabilizing mechanism in the system's evolution. These physics-informed constraints are expected to improve the robustness of the models, particularly in predicting long-term dynamics or extrapolating to unseen scenarios. By integrating such properties, future iterations of I-GILD could bridge the gap between data-driven accuracy and physical consistency, making it a more versatile and reliable tool for a broader range of applications.
}

\bibliographystyle{ieeetr}
\bibliography{./biblio_clean2}
\end{document}